\documentclass[10pt,conference]{IEEEtran}
\IEEEoverridecommandlockouts

\usepackage{cite}
\usepackage{amsmath,amssymb,amsfonts}
\usepackage{algorithmic}
\usepackage{graphicx}
\usepackage{xcolor}
\def\BibTeX{{\rm B\kern-.05em{\sc i\kern-.025em b}\kern-.08em
    T\kern-.1667em\lower.7ex\hbox{E}\kern-.125emX}}
\usepackage{balance}
\usepackage[most]{tcolorbox}
\usepackage{url}
\usepackage{xspace}
\newcommand{\mypara}[1]{\smallskip \noindent\textbf{#1.} \xspace}

\DeclareMathOperator{\cov}{cov}
\DeclareMathOperator{\tr}{tr}

\newcount\DraftStatus  
\definecolor{fixcolor}{rgb}{0.1,0.7,0.3} 
\definecolor{wycolor}{rgb}{0.9,0.1,0.1} 
\definecolor{hycolor}{rgb}{0.7,0.7,0.3} 
\definecolor{zqcolor}{rgb}{0.79, 0.63, 0.86} 

\usepackage{enumitem} 

\begin{document}
\title{How to Select Pre-Trained Code Models for Reuse? A Learning Perspective}

\author{
\IEEEauthorblockN{Zhangqian Bi$^\text{1,2}$, Yao Wan$^\text{1,2,*}$\thanks{*Yao Wan is the corresponding author (wanyao@hust.edu.cn).}, Zhaoyang Chu$^\text{1,2}$, Yufei Hu$^\text{2}$, Junyi Zhang$^\text{2}$, 
Hongyu Zhang$^\text{3}$,\\Guandong Xu$^\text{4}$, Hai Jin$^\text{1,2}$}
\IEEEauthorblockA{$^\text{1}$\textit{National Engineering Research Center for Big Data Technology and System,}}
\IEEEauthorblockA{\textit{Services Computing Technology and System Lab, Cluster and Grid Computing Lab, Wuhan, China}}
\IEEEauthorblockA{$^\text{2}$\textit{School of Computer Science and Technology, Huazhong University of Science and Technology, Wuhan, China}
}
\IEEEauthorblockA{$^\text{3}$\textit{School of Big Data and Software Engineering, Chongqing University, Chongqing, China}
}
\IEEEauthorblockA{$^\text{4}$\textit{School of Computer Science, University of Technology Sydney, Sydney, Australia}
}
}

\maketitle



\begin{abstract}
Pre-training a language model and then fine-tuning it has shown to be an efficient and effective technique for a wide range of code intelligence tasks, such as code generation, code summarization, and vulnerability detection. 
However, pre-training language models on a large-scale code corpus is computationally expensive. 
Fortunately, many off-the-shelf \textit{Pre-trained Code Models} (PCMs), such as CodeBERT, CodeT5, CodeGen, and Code Llama, have been released publicly. These models acquire general code understanding and generation capability during pre-training, which enhances their performance on downstream code intelligence tasks. 
With an increasing number of these public pre-trained models, selecting the most suitable one to reuse for a specific task is essential.
In this paper, we systematically investigate the reusability of PCMs.
We first explore three intuitive model selection methods that select by size, training data, or brute-force fine-tuning. 
Experimental results show that these straightforward techniques either perform poorly or suffer high costs. 
Motivated by these findings, we explore learning-based model selection strategies that utilize pre-trained models without altering their parameters. Specifically, we train proxy models to gauge the performance of pre-trained models, and measure the distribution deviation between a model's latent features and the task's labels, using their closeness as an indicator of model transferability.
We conduct experiments on 100 widely-used open-source PCMs for code intelligence tasks, with sizes ranging from 42.5 million to 3 billion parameters. The results demonstrate that learning-based selection methods reduce selection time to 100 seconds, compared to 2,700 hours with brute-force fine-tuning, with less than 6\% performance degradation across related tasks.
\end{abstract}

\begin{IEEEkeywords}
Model selection, pre-trained code models, model reuse, machine learning.
\end{IEEEkeywords}

\section{Introduction}\label{sec:introduction}
Recently, there has been a surge of \textit{Pre-trained Code Models} (PCMs), such as CodeBERT~\cite{feng2020codebert}, PLBART~\cite{ahmad2021unified}, CodeT5~\cite{wang2021codet5}, CodeGen~\cite{nijkamp2022codegen}, StarCoder~\cite{li2023starcoder}, and Code Llama~\cite{roziere2023code}. 
These models have achieved impressive results in a wide range of code intelligence tasks, including code understanding and code generation. 
In practice, current approaches mostly follow the de-facto \textit{pre-training then fine-tuning} paradigm: pre-training a language model on large-scale code corpus and subsequently fine-tuning it on a specific task~\cite{devlin2019bert,feng2020codebert}.
This approach significantly speeds up the process of developing a deep learning model for a downstream code intelligence task.

However, pre-training a language model on a large-scale code corpus is time-consuming and computationally expensive. 
For instance, Codex, a fine-tuned model based on the GPT-3, costs about $3.1\times 10^6$ GPU hours for pre-training, costing an estimated $4.6\sim 12$ million dollars~\cite{gpt-cost}. 
This underscores the crucial need to leverage existing PCMs whenever possible, a principle that holds particular importance, especially for academia and \textit{small to medium-sized enterprises} (SMEs), which may lack sufficient computing resources.

Fortunately, many off-the-shelf PCMs, such as CodeBERT~\cite{feng2020codebert}, 
PLBART~\cite{ahmad2021unified}, CodeT5~\cite{wang2021codet5}, StarCoder~\cite{li2023starcoder}, and Code Llama~\cite{roziere2023code},
have been released publicly in different online repositories like Hugging Face~\cite{huggingface}, Tensorflow Hub~\cite{tfhub}, ONNX Model Zoo~\cite{onnxmodelzoo}, and PyTorch Hub~\cite{pytorchhub}.
Existing PCMs are typically pre-trained on a specific code corpus using a particular network architecture and validated on specific downstream tasks. 
As a developer, it has become increasingly challenging to choose an appropriate model from the growing number of PCMs for subsequent fine-tuning on a specific task.
An exploratory questionnaire study by~\cite{gong2023intended} finds that developers primarily select models based on their profile metadata, including the task the model performs, the training dataset, and the model's parameter size. 
However, due to the difficulty of understanding and utilizing model parameters~\cite{zhang2021survey}, no existing studies explore model selection that considers the model's weights in software engineering scenarios.

\begin{figure*}[t!]
	\centering
	\includegraphics[width=0.98\textwidth]{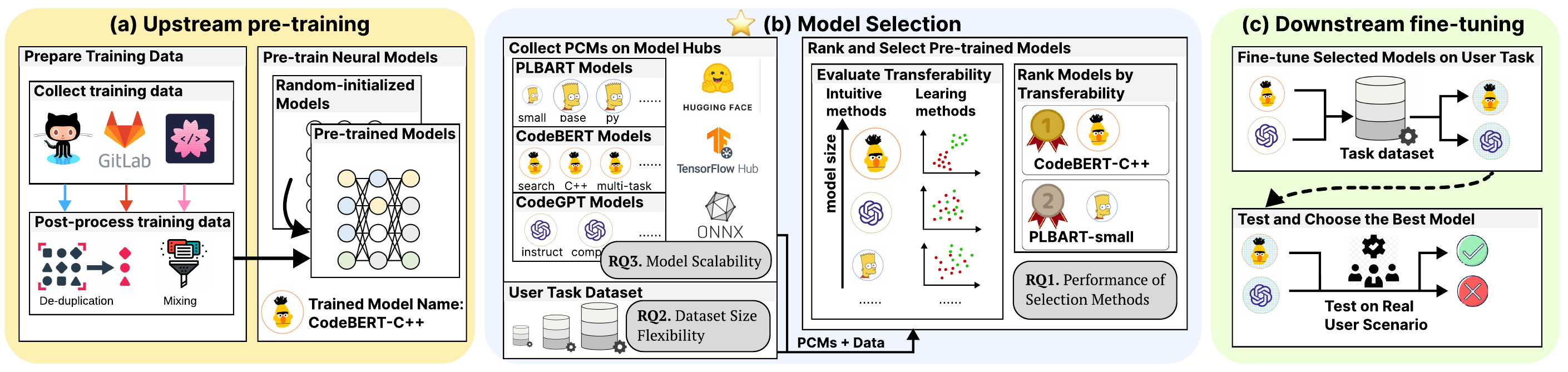}
	\caption{The pipeline of developing and using PCMs, from pre-training to fine-tuning}
	\label{fig_introduction}
	\vspace{-1em}
\end{figure*}

In complementary to recent studies that aim at designing better pre-training techniques for code models, this paper resorts to studying the reusability of pre-trained code models, which has not been systematically explored previously.
Specifically, given a particular task, this paper aims to answer the following question: ``\textit{How to efficiently select a PCM from a model zoo to benefit the target task of interest the most?}''. 
To answer this question, this paper studies the problems from the view of developers when implementing and using PCMs, from pre-training to fine-tuning, as shown in Figure~\ref{fig_introduction}.
Generally, the pipeline of developing and using PCMs can be divided into three stages: 
(1) \textit{Upstream pre-training.} Various models with different architectures are pre-trained on different datasets. (2) \textit{Model selection.} A subset of PCMs is recommended using a ranking strategy. (3) \textit{Downstream fine-tuning.} The selected models are fine-tuned for a specific downstream task.

\mypara{Three Intuitive Approaches and Their Limitations}
As a developer, it is intuitive to select a PCM based on model size, pre-training data, or brute-force fine-tuning. This leads us to derive three intuitive approaches.
\textbf{Approach A:} The first approach hinges on the intuition that the user can straightforwardly choose the PCM of the largest number of parameters possible for deployment. Previous work~\cite{kaplan2020scaling} has illustrated that the learning loss of a pre-trained language model adheres with its size,
which is commonly referred to as the \textit{scaling law}. 
\textbf{Approach B:} Another natural approach is to consider the size of pre-training data of a model, hoping that a larger dataset delivers stronger background knowledge to the task. 
\textbf{Approach C:} An intuitive approach is brute-force fine-tuning, \textit{i.e.,} by fine-tuning all models using the task dataset and subsequently selecting the model that achieves the best performance.

To investigate these three intuitive approaches, we conduct extensive experiments on two distinct downstream tasks:
vulnerability detection, algorithm classification, and programming language identification.
From the results and subsequent analysis of our experiments, we have unveiled several intriguing insights:
a) 
Larger PCMs do not invariably yield improved performance on a particular downstream task.
b) 
Merely selecting models solely based on the size of the pre-training dataset is ineffective, whereas opting for models pre-trained on a multi-lingual code corpus can yield a slight enhancement in performance.
c) 
While the full fine-tuning method is effective, it becomes time-consuming and impractical when dealing with an extensive repository of large pre-trained code models. 
The empirical findings motivate us to learn to efficiently select a pre-trained model with minimal effort.

\mypara{Learning to Select Models}
In this paper, we aim to answer the following research question:
\textit{Can we learn to select the best model from a zoo of PCMs, within a budget (i.e., a limited number of fine-tuning), without the need to fine-tune all models?
}
To answer this question, we investigate three sub-questions focusing on performance (RQ1), adaptability to varying task data sizes (RQ2), and scalability with different numbers of pre-trained models for each selection strategy (RQ3).
We explore five machine learning strategies (i.e., $k$NN~\cite{renggli2022NEEDLE}, Linear Classifier~\cite{yan2020NDS}, SVM~\cite{puigcerver2020scalable}, PARC~\cite{bolya2021PARC}, and HScore~\cite{bao2019HScore}) that learn from parameter for model selection, and establish a benchmark.
The $k$NN, Linear Classifier, and SVM train shallow proxy models on specific tasks, assuming that their accuracy reflects the potential of PCMs after fine-tuning. PARC and HScore, on the other hand, are distribution-based methods that assess the alignment between PCMs' feature distributions and task labels to determine their suitability. 

\begin{table*}[!t]
  \centering
  \caption{Representative PCMs selected from the model zoo, with at least one model sampled from each model type for demonstration. (the \textit{Combined} corpus includes 
  Java and Python functions from GitHub and StackOverflow~\cite{ahmad2021unified}, and \textit{CSN} stands for the CodeSearchNet dataset)
  }

\begin{tabular}{c|lccc}
\hline
\textbf{Model Type} & \textbf{Model Name}                & \textbf{Pre-training Dataset}   & \textbf{Dataset Size} & \textbf{Model Size} \\ \hline
CodeBERT~\cite{feng2020codebert}              & \texttt{codebert-base}      & CodeSearchNet~\cite{husain2019codesearchnet}                   & 6.45M                 & 124.64M             \\
                      & \texttt{graphcodebert-base} & CodeSearchNet~\cite{husain2019codesearchnet}                  & 6.45M                 & 124.64M             \\ \hline
PLBART~\cite{ahmad2021unified}                & \texttt{plbart-base}        & \textit{Combined} & 727M                  & 139.22M             \\
                      & \texttt{plbart-large}       & \textit{Combined}  & 727M                  & 406.02M             \\ \hline
CodeT5~\cite{wang2021codet5}                & \texttt{codet5-small}       & \textit{CSN}+BigQuery          & 8.35M                 & 60.49M              \\
                      & \texttt{codet5-base}       & \textit{CSN}+BigQuery          & 8.35M                 & 222.88M             \\
                      & \texttt{codet5-large}       & \textit{CSN}+BigQuery          & 8.35M                 & 737.63M             \\ \hline
StarCoder~\cite{li2023starcoder}             & \texttt{starcoder-3b}    & The Stack~\cite{kocetkov2022stack}              & 35B                   & 3.0B               \\ \hline
\end{tabular}
    \label{tab:model_statistics}
    \vspace{-1em}
\end{table*}

\mypara{Findings} We conduct comprehensive experiments aimed at comparing the performance of diverse learning strategies for PCM selection. 
The results reveal that learning strategies
can effectively identify the optimal model from 100 opinions.
In contrast to the three intuitive approaches, the adoption of learning-based methods significantly augments the efficiency of the model selection process, reducing the time required from approximately 2,700 hours to a mere 100 seconds. Furthermore, we investigate the impact of budget allocation on the performance of model selection strategies. In order to bolster computational efficiency, we employ a subset of the target task dataset for model selection, where the budget is defined by the size of this subset. Our experiments show that increasing the budget yields superior model selection performance, albeit at the cost of heightened computational expenses. Lastly, we assess the efficacy of these learning strategies when applied to collections featuring varying numbers of models. We employ three collections, each comprising 10, 30, and 100 models, respectively. Our evaluations unequivocally demonstrate that, in all three scenarios, learning strategies outperform their intuitive counterparts.

\mypara{Contributions}
The key contributions of this paper are summarized as follows.
\begin{itemize}[leftmargin=4mm]
\item 
    To the best of our knowledge, we are the first to conduct a systematic study on how to select a model from a zoo of PCMs. Specifically, we explore three intuitive approaches and propose several learning strategies for model selection.
    \item 
    We conduct extensive experiments to demonstrate the effectiveness of learning-based strategies for selecting PCMs in comparison with three intuitive approaches. 
    \item 
    This exploratory study reveals several key findings. First, selecting a PCM based solely on model or pre-training dataset size is unreliable. Second, fine-tuning each PCM brute-force is effective but time- and resource-intensive. Lastly, using learning strategies for model selection proves to be an efficient way to save time and effort.
\end{itemize}

\section{Preliminaries}\label{sec:background}
\subsection{Pre-Trained Code Models (PCMs)}
The PCMs are language models pre-trained on large-scale code corpora using self-supervised techniques. 
They can be generally categorized into three groups by their model structure: encoder-based, decoder-based, and encoder-decoder-based, all utilizing the Transformer as the backbone~\cite{zeng2022extensive}.
Encoder-based models such as CodeBERT~\cite{feng2020codebert} and GraphCodeBERT~\cite{guo2020graphcodebert} aim to predict a masked token utilizing the non-masked words from its bi-directional context.
Decoder-based models such as CodeGPT~\cite{lu2021codexglue}, Code Llama~\cite{roziere2023code} and CodeGeeX~\cite{zheng2023codegeex} aim to predict the next token one by one (\textit{i.e.}, in an \textit{autoregressive} manner) by considering all previous contextual tokens from left to right. 
Encoder-decoder-based models, such as CodeT5~\cite{wang2021codet5}, AlphaCode~\cite{li2022competition} and StarCoder~\cite{li2023starcoder}, jointly train encoder and decoder networks for multiple tasks such as encoding a token as vector embedding or predicting subsequent tokens from the embeddings. 

\subsection{Investigated PCMs}
Without loss of generality, we examine 100 representative PCMs, which are variants of base models including CodeBERT (encoder-based), StarCoder (decoder-based), and PLBART and CodeT5 (encoder-decoder-based).

    \smallskip
    \noindent$\triangleright$ \textbf{CodeBERT~\cite{feng2020codebert}.}
    CodeBERT is an encoder-only pre-trained model for code and natural language modeling. 
    It takes the concatenation of source code and its corresponding natural language comment as input, employing masked language modeling~\cite{devlin2019bert} and replaced token detection~\cite{Clark2020ELECTRA} as training objectives, to predict a masked code or natural language token.
    
    \smallskip
    \noindent$\triangleright$ \textbf{StarCoder~\cite{li2023starcoder}.} StarCoder is a decoder-only model pre-trained on one trillion tokens of over 80 programming languages sourced from GitHub~\cite{kocetkov2022stack}. It employs a fill-in-the-middle pre-training task of predicting a token situated in the middle of a sequence, using tokens from both sides as context. 

    \smallskip
    \noindent$\triangleright$ \textbf{PLBART~\cite{ahmad2021unified}.} Following BART~\cite{lewis2020bart}, a pre-trained model for natural language understanding, PLBART is a PCM with the same architecture and can be used for code understanding tasks. It is pre-trained using the token masking, token deletion, and token infilling objectives.

    \smallskip
    \noindent$\triangleright$ \textbf{CodeT5~\cite{wang2021codet5}.}
    Following the same architecture of T5~\cite{raffel2020exploring}, CodeT5 is an encoder-decoder pre-trained model 
    with an identifier-aware pre-training task that aims to distinguish which code tokens are identifiers. 

Based on these released models, numerous variants have been developed and released from the related communities.
Those model variants are pre-trained in different training datasets (i.e., from multi-lingual code to mono-lingual code), different model sizes (from 42.5M to 3B), or different training configurations (e.g., batch size and learning rate).
All the mentioned PCMs have been hosted in Hugging Face~\cite{huggingface}.
Table~\ref{tab:model_statistics} exemplifies several prevalent PCMs investigated in this paper.
The complete list of PCM variants can be found in the online repository.
\begin{figure*}[!t]
\centering
\includegraphics[width=.98\linewidth]{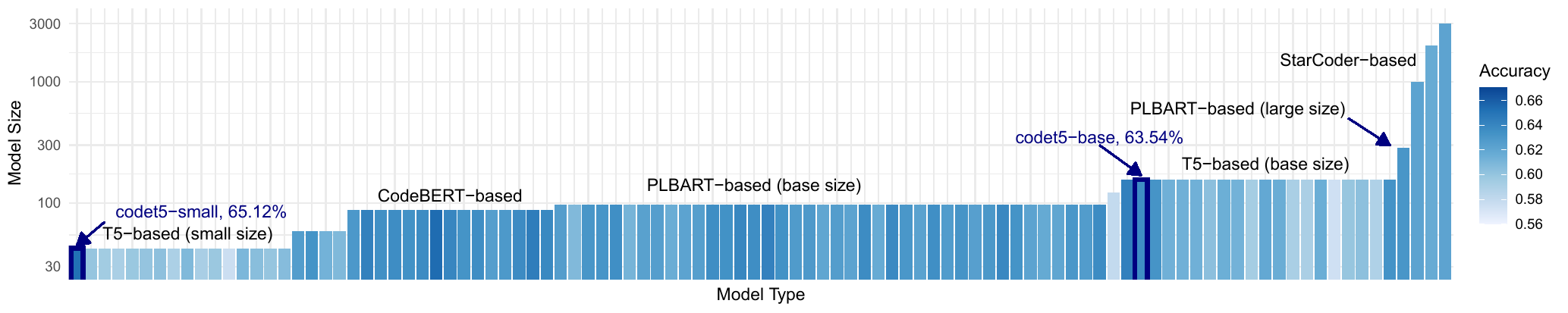}
\caption{
The accuracy of each PCM when adapted to the vulnerability detection task via fine-tuning. 
The accuracy is represented by gradients, with deeper gradients indicating higher values.
Models are sorted in ascending order by size
}
\label{fig:model_size}
\end{figure*}

\subsection{Downstream Tasks of Interest}
The research scope of this paper is narrowed to three representative classification-based code intelligence tasks for simplicity: programming language identification, algorithm classification, and vulnerability detection. These tasks require varying levels of code understanding, ranging from lexical to semantic levels.

    \smallskip
    \noindent$\triangleright$ \textbf{Vulnerability Detection~\cite{zhou2019devign}.} This task aims to determine if the code contains any vulnerabilities. While the output is simple binary values, the detection process itself requires learning comprehensive program semantics to characterize vulnerabilities of high diversity and complexity in source code. 

    \smallskip
    \noindent$\triangleright$ \textbf{Algorithm Classification~\cite{mou2016convolutional}.} This task seeks to categorize the algorithm implemented in given code samples from several categories. This is based on the lexical details such as the naming conventions of the variables and function within the algorithm, and the semantic attributes such as the control and data logic, and flow characteristic of the particular algorithm.
    
    \smallskip
    \noindent$\triangleright$ \textbf{Programming Language Identification~\cite{alreshedy2018scc}.} This task determines the programming language of a given code snippet from a list of possible languages. A good classification relies on the lexical properties (\textit{e.g.}, specific keywords and package names of the programming language) and syntactic properties (\textit{e.g.} a specific form of loop or branch statement) of a language. 

Our analysis focuses on vulnerability detection results, as this challenging task requires a deep understanding of code characteristics and provides the most representative outcomes.
For the other two tasks, we show only the overall performance in Table~\ref{tab:selection_result}.
Learning-based selection methods outperform intuitive methods in all tasks.

\section{Three Task-Agnostic Approaches}
\label{sec:intuitive_exp}

\subsection{
Model Selection Based on Model Size}

We investigate how the size of PCM, i.e., the number of model parameters, influences the effectiveness of vulnerability detection tasks.
The training configuration is referenced in Section~\ref{sec:datasets_and_training}, and it applies to all studies throughout the rest of the paper.

\mypara{Results}
Figure~\ref{fig:model_size} shows the accuracy of each PCM when adapted to the downstream task of vulnerability detection via fine-tuning. The sizes of PCMs have been sorted in ascending order, the smallest model \texttt{codet5-small} has 42.5M parameters, and the largest model \texttt{starcoder-3b} has up to 3B parameters.
From this figure, it is interesting to see that the PCM with a large size does not always yield better performance for the downstream task.
For instance, a specific variant \texttt{Salesforce/codet5-small} with 42.5M parameters attains an accuracy of 65.12\%, whereas a particular variant \texttt{Salesforce/codet5-base}, with 156.57M parameters, achieves only 63.54\% accuracy.
This phenomenon can be attributed to multiple factors including model architecture, training data, and training objectives, necessitating strategic selection instead of simply based on model size.

\begin{tcolorbox}[left=1mm, right=1mm, top=1mm, bottom=1mm]
\textbf{Finding 1.} 
The PCM with a larger size does not always yield better performance for a specific downstream task. 
\end{tcolorbox}

\subsection{
Model Selection Based on Training Data}
We explore the feasibility of model selection based solely on the programming language and the size of the pre-training data.
Due to the high cost of pre-training, we reuse model variants with the same architecture configuration but pre-trained on different programming languages. In previously published works, we find that only the CodeBERT and PLBART models have variants pre-trained on different languages~\cite{zhou2023codebertscore,ahmad2021unified}. 
For the target vulnerability detection task, which uses the C programming language, we select variants pre-trained on individual datasets for C, Java, C++, Python, Go, Ruby, and JavaScript, as well as a variant trained on a combined dataset of these seven languages.
All CodeBERT and PLBART models share exactly the same architecture, size, and training configuration respectively, except for the programming languages of pre-training data.
It is worth noting that the reused models are publicly available pre-trained variants, and the amount of pre-training data, essential for fair comparison, is determined by the original model developers and beyond our control.

\begin{figure}[t!]
\centering
\includegraphics[width=0.98\linewidth]{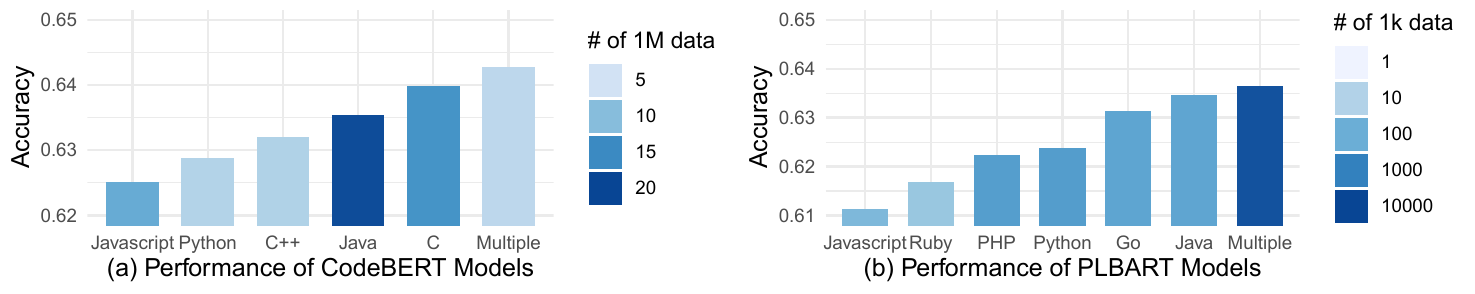}
\caption{
The accuracy of (a) CodeBERT models and (b) PLBART models on the vulnerability detection task. The models are pre-trained on datasets of different sizes and programming languages
}
\label{fig:training_data}
\vspace{-1em}
\end{figure}

\mypara{Results}
Figure~\ref{fig:training_data} shows the results of the selected pre-trained CodeBERT and PLBART models on vulnerability detection. 
From the two sub-tables, we can observe that the models pre-trained on C, C++, Java, and Go, achieve much better performance than that pre-trained on Python and JavaScript, on the task of vulnerability detection.
We attribute it to that C, C++, and Java are more similar to the programming language of the target vulnerability detection dataset (i.e., the C programming language).
It indicates that pre-training models on a dataset of similar programming languages indeed contribute to better performance on the target task.
We also observe that the model pre-trained using a combination of multiple languages has the highest performance in the two tasks.
In Figure~\ref{fig:training_data}(a), even the dataset of multiple languages (sourced from the CodeSearchNet~\cite{husain2019codesearchnet} dataset) is significantly smaller compared to a single-language dataset (sourced from the CodeParrot~\cite{codeparrot} dataset, which is a clean copy from GitHub), the CodeBERT model trained on this multilingual dataset still surpasses those trained on a single language. This indicates that including multiple languages in training benefits the understanding of the target language's tasks, analogous to existing study~\cite{DBLP:journals/corr/abs-2204-09653}.
%
\begin{tcolorbox}[left=1mm, right=1mm, top=1mm, bottom=1mm]
\textbf{Finding 2.} 
For PCMs that share the same architectures (CodeBERT), we recommend selecting the model that is pre-trained on a dataset of multiple or similar programming languages. 
However, we cannot simply select the model pre-trained on a larger dataset.
\end{tcolorbox}

\begin{figure}[t!]
	\centering
    \includegraphics[width=0.48\textwidth]{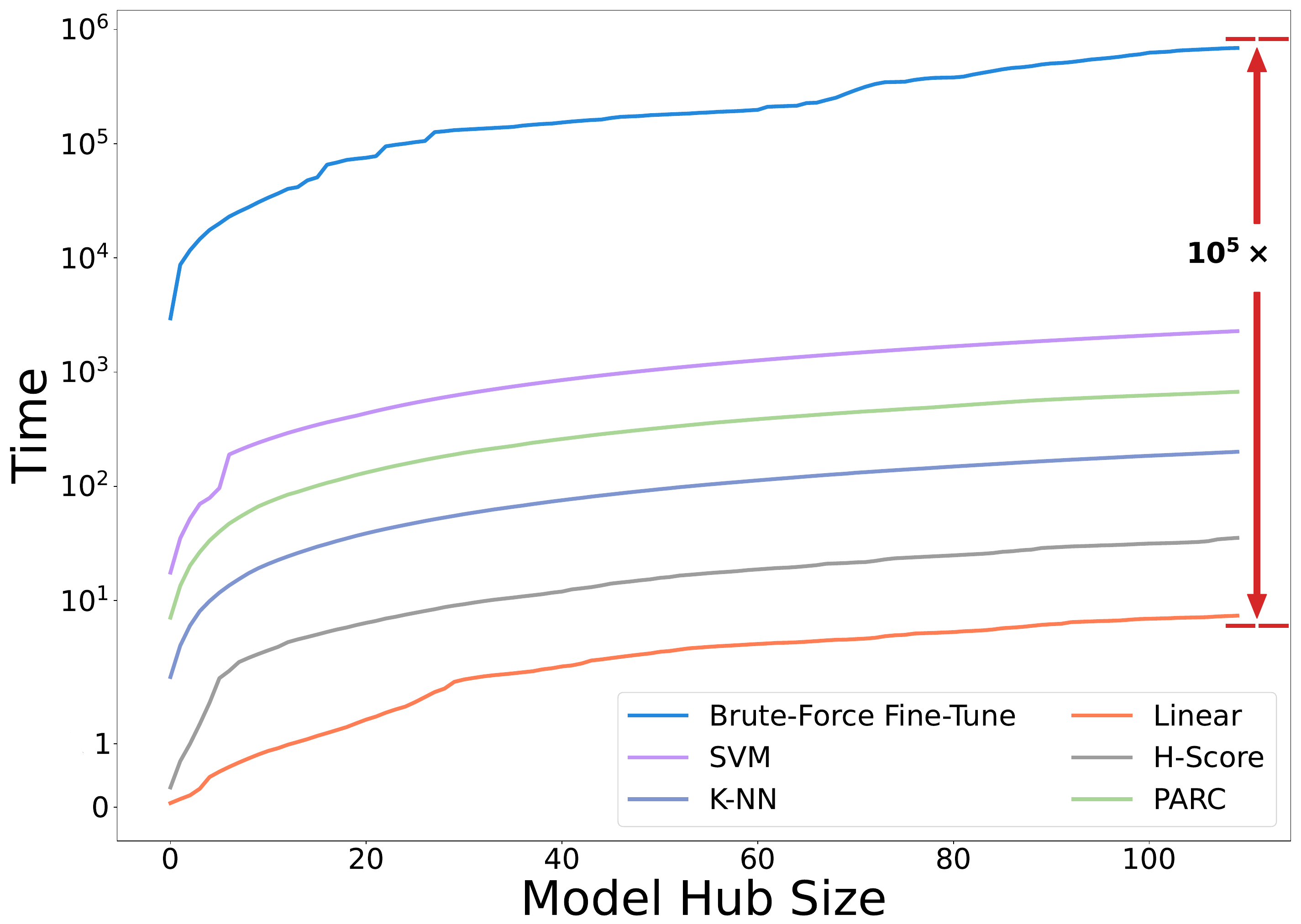}
    \caption{The time cost of brute-force fine-tuning and various learning strategies for model selection. The horizontal axis (Model Hub Size) represents the number of models involved in the selection process (measured per model), and the vertical axis (Time) indicates the selection time cost in seconds
    }
    \vspace{-1em}
    \label{fig:time}
\end{figure}

\subsection{Model Selection Based on Brute-Force Fine-Tuning}
\label{sec_fine_tuning}
With unlimited computational resources, one straightforward approach to model selection is to exhaustively fine-tune each candidate and choose the one that yields the optimal results. This method is commonly referred to as \textit{brute-force fine-tuning}. In this study, we investigate the practicality of brute-force fine-tuning for selecting the best PCM. The implementation follows Approaches A and B, with time costs for fine-tuning each model recorded.

\mypara{Results}
Figure~\ref{fig:time} depicts the time cost (seconds in log-scale) of brute-force fine-tuning when increasing the size of the PCM zoo.
Note that, in this figure, we also record the time cost of several learning strategies (e.g., $k$-NN and PARC) for model selection that will be introduced in {Sec.~\ref{sec:methodology}}.
From this figure, we can observe that brute-force fine-tuning costs much more time than the learning strategies, up to $10^5\times$ as the size of PCMs increases to 100.
Furthermore, we can see that the brute-force fine-tuning approach costs up to $10e\mathrm{+}6$ seconds ($\sim$ 2,700 hours) and the time cost will increase linearly with the size of the PCM zoo, which is unacceptable in practice.
\begin{tcolorbox}[left=1mm, right=1mm, top=1mm, bottom=1mm]
\textbf{Finding 3.} 
Even though brute-force fine-tuning each candidate PCM and selecting the one with the best performance is effective, it is time-consuming and unacceptable as the size of the PCM zoo increases. 
\end{tcolorbox}

\section{Learning to Select Models}
\label{sec:methodology}

\begin{figure*}[t!]
    \centering
    \includegraphics[width=\textwidth]{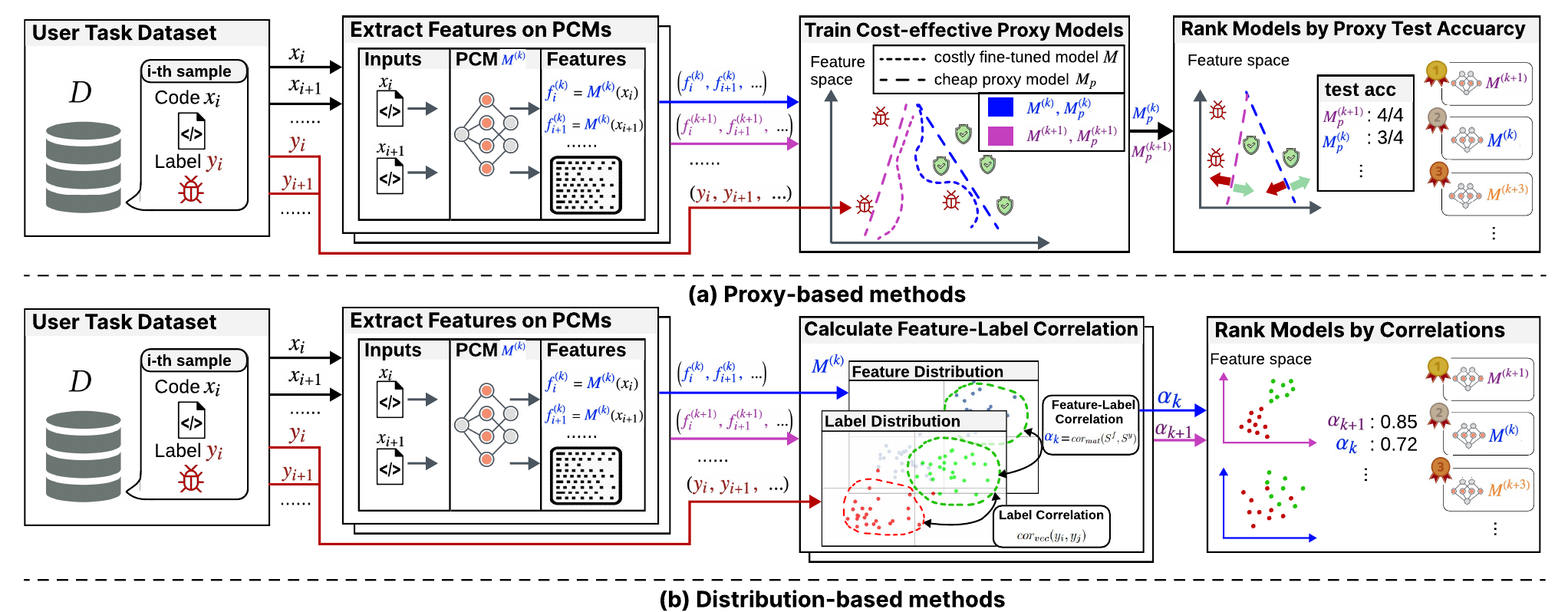}
    \vspace{-1em}
    \caption{An illustration of the (a) proxy-based and (b) distribution-based model selection strategies
    }
        \vspace{-1em}
    \label{fig:model_selection_methods}
\end{figure*}

The aforementioned empirical findings motivate us to learn to efficiently select a pre-trained model with minimal effort.
In this section, we introduce two types of learning strategies (i.e., proxy-based and distribution-based) for PCM selection.
Figure~\ref{fig:model_selection_methods} shows an overview of the investigated learning methods for model selection.

\subsection{Problem Formulation}
The goal of model selection is to find a source PCM $M$ from a collection of $n$ models $\mathbb{M}=\{M_1, M_2,\ldots,M_n\}$, which have a high transferability score to a target task $t \in T$.
We interpret the PCM selection as a ranking problem, where models are ranked based on their transferability scores, and those with higher scores are selected.
We start by defining the transferability score and the computing budget.

\mypara{Transferability Score}
Given a specific pre-trained model $M$, a task $t$, the \textit{transferability score} of model $M$ with respect to task $t$, denoted as $\alpha(M, t)$, quantifies the expected performance of $M$ when applied to task $t$.
An intuitive way to determine the transferability score is by performing brute-force fine-tuning for the downstream task and using the resulting accuracies as the scores.
However, this approach demands substantial computing resources. 
Thus, we introduce a \textit{computing budget} constraint to limit the resources allocated for fine-tuning the models. 

\mypara{Computing Budget}
The \textit{computing budget}, denoted as $b$, represents the number of top-performing models selected for fine-tuning on the target task.
Intuitively, the user can only select a small set of models for fine-tuning, usually the top 1 or top 5 model, this is limited by the computing budget.
Given the dataset $D = (X, Y)$ of target task $t$, the computing budget $b$ and the collection of models $\mathbb M$,
a selection approach $A$ computes a transferability score $\alpha_i(M_i, t)$ for each model $M_i \in \mathbb M$, by learning from the model parameters and the task dataset $D$.
After all the scores are computed, the models are sorted by corresponding $\alpha_i$ to generate a ranked list $\{M^{\alpha}_{i}\}_{i=1}^n$, and the $b$ best-performing models are chosen for further fine-tuning.
In particular, for the intuitive way of selecting models by performing brute-force fine-tuning for the downstream task,
we refer to this method as $A^*$, and the true ranked list generated by this method as $\{M^{\alpha^*}\}_{i=1}^n$.

The quality of the selection method $A$ is assessed by comparing its selected top-$b$ models to the best top-$b$ models using an evaluation metric $\mu$, which can be expressed as:
\begin{equation}
     \mu(A) = \mu(\{M_i^{\alpha}\}_{i=1}^{b}, \{M_i^{\alpha^*}\}_{i=1}^b)
\end{equation}

\subsection{Proxy-Based Methods}
\label{sec:proxy_based_methods}

A suitable PCM for the task produces high-quality latent representations of the input sequences.
We utilize this insight by leveraging the quality of latent representations to assess the performance of the corresponding PCMs.
The \textit{proxy-based method},
as illustrated in Figure~\ref{fig:model_selection_methods} (a), 
learns the transferability score of a PCM by training \textit{proxy} models using PCM-generated features and task labels.
Features are the vector representations of the input sequences.
A PCM $M:X \to F$ maps a sequence from input code space $X$ to its feature space $F$ 
through a forward pass, building a feature-label dataset $D^{(k)}$ given by:
\begin{equation}
    D^{(k)} = \{(f^{(k)}_i, y_i)\ |\ f^{(k)}_i = M^{(k)}(x_i), (x_i,y_i)\in D\}
\end{equation}
where $M^{(k)}$ represents the $k$-th PCM in the model zoo, $x_i, y_i$ represents the input sequences with its label, and $f^{(k)}_i$ represents the features generated by $M^{(k)}$ for $i$-th input sequence in $D$.
After obtaining $D^{(k)} = (F, Y)$, this method trains and evaluates a proxy model $M_p$ for each PCM $M$ in the zoo using a particular training method $\mathcal T$. The process can be formulated as follows:
\begin{equation}
    X \to F \to \mathcal T(M_p, F, Y) \leftarrow Y
\end{equation}
The test accuracy of this proxy model $M_p$ then serves as the transferability score to determine how well the PCM $M$ performs on the target task.
Different training methods $\mathcal{ T}$ is widely used in several computer vision papers~\cite{puigcerver2020scalable, yan2020NDS, renggli2022NEEDLE}, we follow these works to use $k$-NN~\cite{renggli2022NEEDLE}, linear~\cite{yan2020NDS}, and SVM proxy classifiers~\cite{puigcerver2020scalable} for this study, motivated by their interpretability, simplicity, and effectiveness demonstrated in prior studies. 
In our implementation, for the $k$-NN method, we use different values $k$ of $1$, $3$, and $5$, 
for the linear classifier method, we use a single-layer linear neural network,
and for the SVM classifier, we set the regularization parameter $C = 1.0$.

\subsection{Distribution-Based Methods}
A PCM $M$ is suitable for the task when two code features that are proximate in the PCM's feature space share similar or identical labels in the task's label space.
As illustrated in Figure~\ref{fig:model_selection_methods}(b), the distribution-based method learns the transferability score by calculating the 
correlation
between feature distribution generated by the PCM and label distribution from the task.
For a feature-label dataset $D^{(k)} = \{(f^{(k)}, y)\}$, this method first computes two matrices $S^f$ and $S^y$ to represent the distributions of features and labels, respectively:
\begin{align}
    S^f = [s_{ij}^f]_{n\times n},\ \ &\text{where}\ \ s_{ij}^f = cor_{vec}(f_i, f_j)\, \\
    S^y = [s_{ij}^y]_{n\times n},\ \ &\text{where}\ \ s_{ij}^y = cor_{vec}(y_i, y_j)\,
\end{align}
where $f_i, f_j \in F$ are the model-extracted features, $y_i, y_j \in Y$ are the task labels, $cor$ serves as a vector correlation 
measurement metric. 
The correlation of these two distributions then serves as the transferability score:
\begin{equation}
    \alpha(M^{(k)}, t) = cor_{mat}(S^f, S^y)
\end{equation}
In this study, we utilize two distribution-based methods, namely the PARC (\textit{Pairwise Annotation Representation Comparison})~\cite{bolya2021PARC} and the H-Score~\cite{bao2019HScore}, they differ mainly in the definition of the correlation metrics $cor$.

The PARC approach~\cite{bolya2021PARC} defines the two correlation metrics as follows:
\begin{align}
    &cor_{vec}(u, v) = 1 - Pearson(u,v) \\
    &cor_{mat}(S^f, S^y) = Spearman(S^f, S^y)\, 
\end{align}
where $Pearson$ and $Spearman$ represent the Pearson and Spearman correlation coefficient, respectively, and $u, v$ can be features and labels (treated as one-dimension vectors). 
The idea of using dissimilarities $1-Pearson(u, v)$ to represent the feature and label space is that dissimilarity can enhance the contrast between the two features, making it easier to differentiate between them.


The H-Score~\cite{bao2019HScore} defines different $cor$ operators for features and labels, as follows:
\begin{align}
    s_{ij}^f = cor_{vec}(f_i, f_j) &= \cov(f_i, f_j) \\ 
 s_{ij}^y = cor_{vec}(y_i, y_j) &= \cov(\mathbb E[f | y_i], \mathbb E[f | y_j]) 
\end{align}
where $f_i, f_j \in F$ and $y_i, y_j \in Y$.
Here, $S^f = [s_{ij}^f]_{n\times n}$ is known as \textit{feature redundancy} matrix in the context of H-Score because the larger magnitude means that the features are correlated more closely, thus containing more redundant information, and $S^y = [s_{ij}^y]_{n\times n}$ is referred to as the \textit{inter-class variance}, as it measures the dissimilarities between features when their corresponding labels are different.

Finally, the H-Score defines the $corr$ operator as follows:
\begin{equation}
    cor_{mat}(S^f, S^y) = \tr((S^f)^{-1}S^y)\,
\end{equation}
aiming to simultaneously maximize the inter-class variance and minimize the feature redundancy.

\section{Learning Strategies Evaluation}
To validate the effectiveness of our introduced learning strategies for model selection, this section aims to answer the following \textit{Research Questions} (RQs).
\begin{itemize}[leftmargin=4mm]
    \item \textbf{RQ1.} \textit{To what extent can we learn to select the best model from a diverse collection of PCMs without brute-force fine-tuning?}
    We conduct extensive experiments on a large and diverse collection containing 100 model variants from 9 different architectures, to explore whether we can learn to select the best model without brute-force fine-tuning. 
    \item \textbf{RQ2.} \textit{Can we adapt learning strategies to accommodate different dataset sizes?}
    We adapt model selection strategies within different computing budgets to examine if learning strategies remain effective when computing resources are constrained. We limit the budget size from 1,000 to 5,000 to observe if learning methods are adaptable to varying computing resources.
    \item \textbf{RQ3.}  \textit{Can we effectively select a model as the number of available models scales?}
    We investigate the effectiveness of model selection across different scales of model collections. 
    We use three model collections with 10, 30, and 100 models, to investigate if the learning strategies can select the high-performing models from collections of varying sizes.    
\end{itemize}

\subsection{Datasets and Training Setup}
\label{sec:datasets_and_training}
We employ three code intelligence tasks with their respective datasets for this benchmark: programming language identification, algorithm classification, and vulnerability detection.
For the vulnerability detection task, we adopt the \textit{Devign} dataset~\cite{zhou2019devign}.
\textit{Devign} includes 27,318 manually labeled functions sampled from two large open-source projects (\textit{i.e.}, QEMU and FFmpeg) written in C. 
The dataset is created by collecting security-related commits and extracting vulnerable or non-vulnerable functions from the labeled commits, and the task is formulated as a binary classification to predict the presence (1) or absence (0) of a vulnerability.
We fine-tune the models following the training configuration in a widely acknowledged benchmark paper CodeXGLUE~\cite{lu2021codexglue}. All models are fine-tuned for 5 epochs, with a learning rate of $2e\mathrm{-}5$, and a batch size of 8.
For the algorithm classification task, we use the \textit{POJ-104} dataset~\cite{mou2016convolutional}, as this task is previously investigated by Peng et al.~\cite{peng2021could}.
This dataset is designed to classify source code into one of 104 distinct algorithm categories. It comprises 52,000 programs, each representing one of these categories, collected from an \textit{online judge} (OJ) system.
Each model is fine-tuned using a learning rate of $2e\mathrm -5$ and a batch size of 32 for 5 epochs, following the same configurations as in Lu et al.~\cite{lu2021codexglue}.
For the programming language identification task, we use the \textit{SCC} dataset\cite{alreshedy2018scc}.
This dataset contains 237,803 code fragments extracted from Stack Overflow posts with language labels. The dataset consists of 21 different programming languages: Bash, C, C\#, C++, CSS, Haskell, HTML, Java, JavaScript, Lua, Objective-C, Perl, PHP, Python, R, Ruby, Scala, SQL, Swift, VB, and Markdown. 
Since the training configuration in SCC~\cite{alreshedy2018scc} does not involve neural networks, we fine-tuned each model using a learning rate of $2e\mathrm -5$ and a batch size of 32 for 5 epochs, consistent with the other two tasks. Convergence was observed after 2 epochs.
Each fine-tuned model is evaluated after each training epoch on a validation dataset, and the model that achieves the best performance on the validation dataset is dumped to test.

\subsection{Evaluation Metrics}
\label{sec:metrics}
We use two performance metrics (i.e., $\NDCG@k$ and Top-$k$ Relative Accuracy) and an efficiency metric (i.e., Time Cost) to measure the quality of a selection method. Here, $k$ represents the computing budget, \textit{i.e.}, the number of models a user can select for further fine-tuning.
These metrics are commonly adopted in existing model selection literature within the machine learning community.

\mypara{Normalized Discounted Cumulative Gain ($\NDCG@k$)~\cite{jarvelin2002cumulated}}
By considering model selection as a ranking problem, we adapt information retrieval metrics to evaluate the selection quality.
Normalized Discounted Cumulative Gain is an evaluation metric that quantifies the quality of the retrieved model list. 
The \textit{Discounted Cumulative Gain (DCG)} accumulated at a rank position $k$ in a ranking list $R$ is defined by:
\begin{equation}
\small
  \DCG@k(R) = \sum_{i=1}^{k}\frac{2^{r(i)}-1}{\log_2(i+1)}\,
\end{equation}
where $r(i)$ is the performance of model rank $i$ trained on target task. The NDCG is a normalized version of the DCG of the predicted ranking, defined by:
\begin{equation}
\small
  \NDCG@k(R)=\frac{\DCG@k(R_{\text{pred}})}{\DCG@k(R_{\text{true}})}\,
\end{equation}
where $R_{\text{pred}}$ and $R_{\text{true}}$ denotes the predicted and ground-truth ranking list, respectively.
A high value $NDCG@k$ indicates that the selected top-$k$ models align closely with the ideal top-$k$ models, thus the selection method has high precision.

\mypara{Top-K Relative Accuracy ($Rel@k$)~\cite{li2021NLEEP,agostinelli2022STABLE}}
Finding the best-performing model exactly from a large zoo of models can be an unrealistic goal with the growing number of PCMs.
If a method $A$ chooses high-performing models in its top-$k$ selection, even if it does not include the highest-performing one, it should not be overly penalized.

This metric computes the performance discrepancy between the best-performing model out of $k$ selected models and the best-performing model within the model zoo.
It is defined by:
\begin{equation}
    Rel@k(\{M^{\alpha}_i\}_{i=1}^n)=\frac{\max_{M_i\in\{M^{\alpha}\}_{i=1}^k}Acc(M_i)}{Acc(M_1^{\alpha^*})}\,
\end{equation}
where $\{M^{\alpha}\}_{i=1}^k$ is the ranked list of selected top-$k$ models, and $M^{\alpha^*}_1$ is the best-performing model in the zoo.
A higher $Rel@k$ value indicates that the selected $k$ models are closely aligned with the actual best model.
Consequently, fine-tuning this subset of top models and then selecting the most effective one is likely to yield strong performance. 
If $Rel@k=1$, the best model is within the selected $k$ models.
We normalize the $Rel@k$ score by subtracting the accuracy of the model with the lowest performance, ensuring the values fall inclusively between 0 and 1.

\mypara{Time Cost}
In addition to the precision of each model selection method, we also examine the execution time of each method to measure their efficiency.
All experiments are conducted in a Linux server, with two Intel Xeon Gold 5117 CPUs and 4 Tesla V100 GPUs. 
\begin{table}[!t]
\centering
\caption{
Rel$@k$ and NDCG$@k$ of four types of model selection methods on the vulnerability detection, algorithm classification, and programming language identification task
}
\setlength{\tabcolsep}{4pt} 
\resizebox{0.98\linewidth}{!}{
\begin{tabular}{lcccccc}
\hline
\textbf{Metric}             & \multicolumn{3}{c}{NDCG@$k$}                                             & \multicolumn{3}{c}{Rel@$k$}                                              \\ \hline
\textbf{$k$}                  & \multicolumn{1}{c}{1} & \multicolumn{1}{c}{5} & \multicolumn{1}{c}{10} & \multicolumn{1}{c}{1} & \multicolumn{1}{c}{5} & \multicolumn{1}{c}{10} \\ \hline
\multicolumn{7}{c}{\textit{Vulnerability Detection}}\\
\hline
\textbf{Proxy-based}        & 0.42                  & 0.53                  & \textbf{0.59}                    & 0.35                  & \textbf{0.71}                  & \textbf{0.94}                   \\
\textbf{Distribution-based} & 0.39                  & \textbf{0.56}                  & 0.58                    & 0.30                   & 0.63                  & 0.73                   \\
\textbf{Model size}         & \textbf{0.45}                  & 0.49                  & 0.54                    & \textbf{0.38}                  & 0.44                  & 0.55                   \\
\textbf{Dataset size}       & 0.30                   & 0.34                  & 0.38                    & 0.23                  & 0.34                  & 0.39                   \\ \hline
\multicolumn{7}{c}{\textit{Algorithm Classification}}\\
\hline
\textbf{Proxy-based}        & 0.74                  & 0.78                  & 0.83                    & 0.96                  & 0.98                  & 0.98                   \\
\textbf{Distribution-based} & \textbf{0.82}                  & \textbf{0.86}                  & \textbf{0.93}                    & \textbf{0.98}                  & \textbf{1.00}                     & \textbf{1.00}                      \\
\textbf{Model size}         & 0.33                  & 0.77                  & 0.81                    & 0.85                  & 0.97                  & 1.00                      \\
\textbf{Dataset size}       & 0.56                  & 0.69                  & 0.69                    & 0.92                  & 0.98                  & 0.99                   \\ \hline
\multicolumn{7}{c}{\textit{Programming Language Identification}} \\
\hline
\textbf{Proxy-based}        & 0.78                  & 0.75                  & 0.78                    & 0.79                  & \textbf{0.93}                  & 0.97                   \\
\textbf{Distribution-based} & \textbf{0.79}                  & \textbf{0.87}                  & \textbf{0.88}                    & \textbf{0.87}                  & 0.92                  & 0.94                   \\
\textbf{Model size}         & 0.44                  & 0.49                   & 0.59                    & 0.23                  & 0.58                  & 0.91                   \\
\textbf{Dataset size}       & 0.44                  & 0.50                   & 0.61                    & 0.42                  & 0.60                   & \textbf{1.00}                      \\ \hline
\end{tabular}
}
\label{tab:selection_result}
\vspace{-1em}
\end{table}

\begin{figure}[t]
    \centering
    \includegraphics[width=0.98\linewidth]{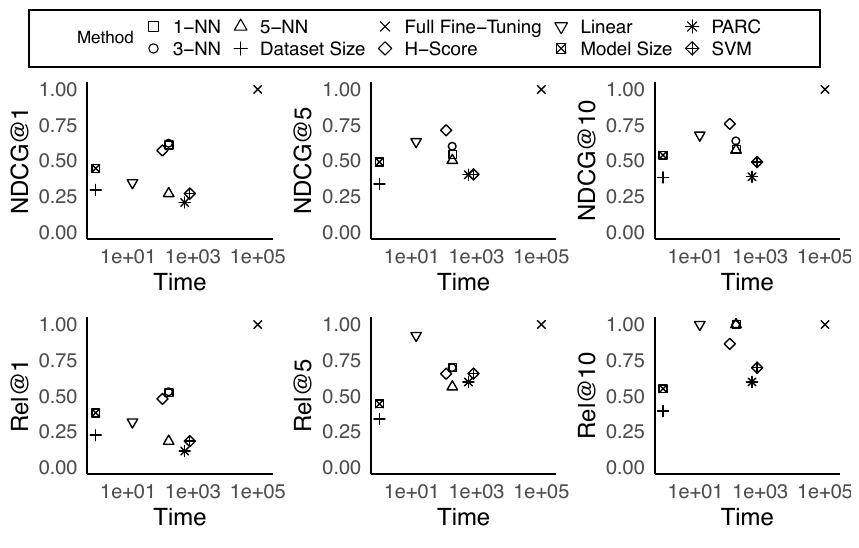}
    \caption{NDCG Scores, Rel$@k$ Accuracies, and Time Cost of model selection strategies on the vulnerability detection tasks}    
    \label{fig:selection_result}
    \vspace{-1.5em}
\end{figure}

\subsection{Overall Performance (RQ1)} 
\label{sec:rq1}
Table~\ref{tab:selection_result} shows the $\NDCG@k$ and the $\Rel@k$ scores of each model selection strategy on the \textsf{Devign}, \textsf{POJ-104}, and the \textsf{SCC} dataset.
We sample 1000 points uniformly from each dataset for model selection and run the sampling and selection 5 times to reduce sampling bias.
The transferability score is calculated as the average of these individual run scores.
As shown in this table, the learning-based selection methods surpass simply selecting the models based on their size, or the size of the training dataset.

We take a closer look focusing on the vulnerability detection task, the results are presented in Figure~\ref{fig:selection_result}.
As shown in the figure, the Brute-Force Fine-Tuning method which serves as the gold selection method has the highest quality score ($\NDCG@k = 1.0, \Rel@k = 1.0$), whereas costing more than $1e$+5 seconds to draw the selection. 
Two intuitive selection methods based on model size or dataset side do not require extra inference or fine-tuning, shown in the leftmost in each figure with time $t=0$. They do not provide good performance, which is analogous to the observation in Section~\ref{sec:intuitive_exp}.
We attribute to it that these methods cannot learn from the task dataset, necessitating learning strategies for model selection.

For learning methods, when they are only allowed to select one model, their performance seems random, with some even underperforming intuitive baselines.
However, increasing the computing budget to 5 and 10 models consistently favors learning methods over intuitive ones,
and the time cost of learning-based methods is over 100 times less than fine-tuning all the models.

\begin{tcolorbox}[left=1mm, right=1mm, top=1mm, bottom=1mm]
\textbf{Answer to RQ1.} 
It is possible to select high-performing models from a zoo of PCMs without fine-tuning.
Selecting through learning-based methods is more effective than 
selecting models based on size or pre-training dataset. 
\end{tcolorbox}

\begin{table*}[!t]
\centering
\caption{
Rel@$k$ and NDCG@$k$ of each model selection strategy with varying numbers of candidate models
}
\setlength{\tabcolsep}{5.6pt} 
\begin{tabular}{l|cccccc|cccccc|cccccc}
\hline
                      & \multicolumn{6}{c|}{\textit{Number of Models = 10}}                                  & \multicolumn{6}{c|}{\textit{Number of Models = 30}}                                  & \multicolumn{6}{c}{\textit{Number of Models = 100}}                                           \\ \hline
\textbf{Metric}       & \multicolumn{3}{c}{Rel@$k$}            & \multicolumn{3}{c|}{NDCG@$k$}                   & \multicolumn{3}{c}{Rel@$k$}            & \multicolumn{3}{c|}{NDCG@$k$}                   & \multicolumn{3}{c}{Rel@$k$}                     & \multicolumn{3}{c}{NDCG@$k$}                    \\ \hline
\textbf{$k$}       & 1             & 5             & 10   & 1             & 5             & 10            & 1    & 5             & 10            & 1             & 5             & 10            & 1             & 5             & 10            & 1             & 5             & 10            \\ \hline
\textbf{1-NN}         & 0.63          & \textbf{1.00} & 1.00 & 0.68          & 0.70          & 0.84          & 0.52 & \textbf{1.00} & \textbf{1.00} & 0.67          & 0.59          & 0.65          & 0.52          & 0.70          & \textbf{1.00} & 0.61          & 0.54          & 0.58          \\
\textbf{3-NN}         & \textbf{1.00} & \textbf{1.00} & 1.00 & 0.26          & 0.69          & 0.84          & 0.70 & 0.92          & 0.92          & 0.26          & 0.58          & 0.71          & \textbf{0.53} & 0.69          & \textbf{1.00} & \textbf{0.62} & 0.60          & 0.64          \\
\textbf{5-NN}         & 0.63          & \textbf{1.00} & 1.00 & 0.68          & 0.69          & 0.84          & 0.52 & \textbf{1.00} & \textbf{1.00} & 0.67          & 0.63          & 0.72          & 0.18          & 0.56          & \textbf{1.00} & 0.27          & 0.50          & 0.57          \\
\textbf{Linear}       & \textbf{1.00} & \textbf{1.00} & 1.00 & 0.26          & 0.66          & 0.85          & 0.70 & \textbf{1.00} & \textbf{1.00} & 0.26          & 0.52          & 0.57          & 0.32          & \textbf{0.92} & \textbf{1.00} & 0.35          & 0.63          & 0.68          \\
\textbf{SVM}          & 0.88          & \textbf{1.00} & 1.00 & \textbf{0.91} & 0.71          & 0.88          & 0.92 & 0.92          & 0.93          & 0.90           & 0.65          & 0.74          & 0.18          & 0.65          & 0.70           & 0.27          & 0.41          & 0.49          \\
\textbf{H-Score}      & 0.63          & 0.98          & 1.00 & 0.68          & \textbf{0.90} & \textbf{0.95} & 0.40 & 0.64          & \textbf{1.00} & \textbf{0.94} & \textbf{0.86} & 0.81          & 0.48          & 0.66          & 0.86          & 0.57          & \textbf{0.71} & \textbf{0.76} \\
\textbf{PARC}         & 0.73          & 0.88          & 1.00 & 0.78          & 0.81          & 0.91          & 0.56 & 0.56          & 0.69          & 0.77          & 0.77          & 0.77          & 0.11          & 0.59          & 0.59          & 0.21          & 0.40           & 0.39          \\ \hline
\textbf{Model Size}   & 0.63          & \textbf{1.00} & 1.00 & 0.68          & 0.65          & 0.83          & 0.27 & 0.35          & 0.39          & 0.73          & 0.83          & \textbf{0.84} & 0.38          & 0.44          & 0.55          & 0.45          & 0.49          & 0.54          \\
\textbf{Dataset Size} & 0.63          & \textbf{1.00} & 1.00 & 0.68          & 0.72          & 0.89          & 0.06 & 0.21          & 0.29          & 0.61          & 0.68          & 0.74          & 0.23          & 0.33          & 0.39          & 0.30           & 0.34          & 0.38          \\ \hline
\end{tabular}
\label{tab:number_of_models}
\end{table*}

\begin{figure}[t!]
	\centering
        \includegraphics[width=0.8\linewidth]{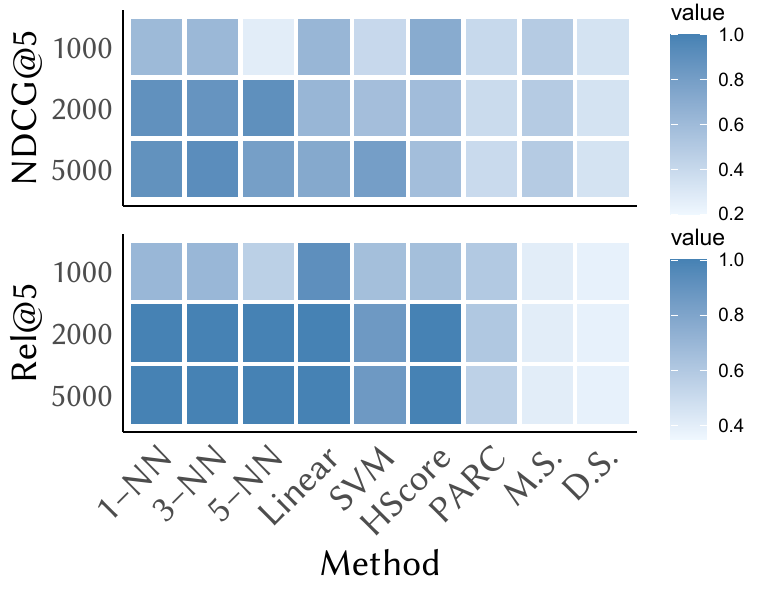}
	\label{fig:budget_trend_a}
 \vspace{-0.5em}
        \caption{NDCG@5 and Rel@5 heatmap of each learning-based method for model selection with varying budget size (M.S. stands for Model Size, and D.S. stands for Datset Size)}
	\label{fig:budget_plot}
	\vspace{-1em}
\end{figure}

\subsection{Adaptability (RQ2)}
In this RQ, we assess whether these learning strategies can adapt to different sizes of task datasets.
Users typically cannot supply the complete dataset for model training and are only able to provide a very limited number of probe samples at the stage of model selection.
To stimulate this scenario, we measure the NDCG and top-$k$ relative accuracy of model selection methods for the vulnerability detection task.
Here, the user provides 1000, 2000, and 5000 probe samples to select a high-performing model.
The samples are uniformly drawn from the dataset and the experiment is repeated 5 times to reduce sampling bias.

\mypara{Results}
Figure~\ref{fig:budget_plot} illustrates the performance of each learning-based model selection strategy with varying numbers of probe samples. As depicted in the figure, the learning-based methods consistently outperform the models selected without considering the user dataset, such as the Model Size and Data Size methods. When the sample size increases to 2,000 and 5,000, all learning strategies surpass these two intuitive approaches.

\begin{tcolorbox}[left=1mm, right=1mm, top=1mm, bottom=1mm]
\textbf{Answer to RQ2.} 
In summary, learning strategies outperform intuitive methods with only a dataset of 1,000 user samples. Furthermore, their performance consistently improves as the sample size increases.
\end{tcolorbox}

\subsection{Scalability (RQ3)}
In this RQ, we investigate whether the learning strategies remain effective when the number of candidate models grows.
Our model repository is up-to-date as of September 2023. We have selected the latest 10, 30, and 100 models based on their release dates.

\mypara{Results}
Table~\ref{tab:number_of_models} demonstrates the NDCG and top-$k$ relative accuracy of each selection method for the vulnerability detection task.
The table shows that the learning strategies can consistently select the high-performing models among various numbers of candidate models, whereas simply selecting based on model size or data size exhibits unstable performance.
Furthermore, consider the points where $\Rel@k=1.0$ (highlighted in bold), which indicates that the top-$k$ selection includes the best-performing model. 
The learning-based selection methods consistently identify the best-performing model, accurately pinpointing it among the top 10 out of 100 models.
However, selections based on model size and dataset size exhibit a random trend as the number of candidate models increases.
\begin{tcolorbox}[left=1mm, right=1mm, top=1mm, bottom=1mm]
\textbf{Answer to RQ3.}
Learning strategies are effective as model collection scales, exhibiting stable selection accuracy.
\end{tcolorbox}

\begin{figure}[t!]
	\centering
    \includegraphics[width=0.8\linewidth]{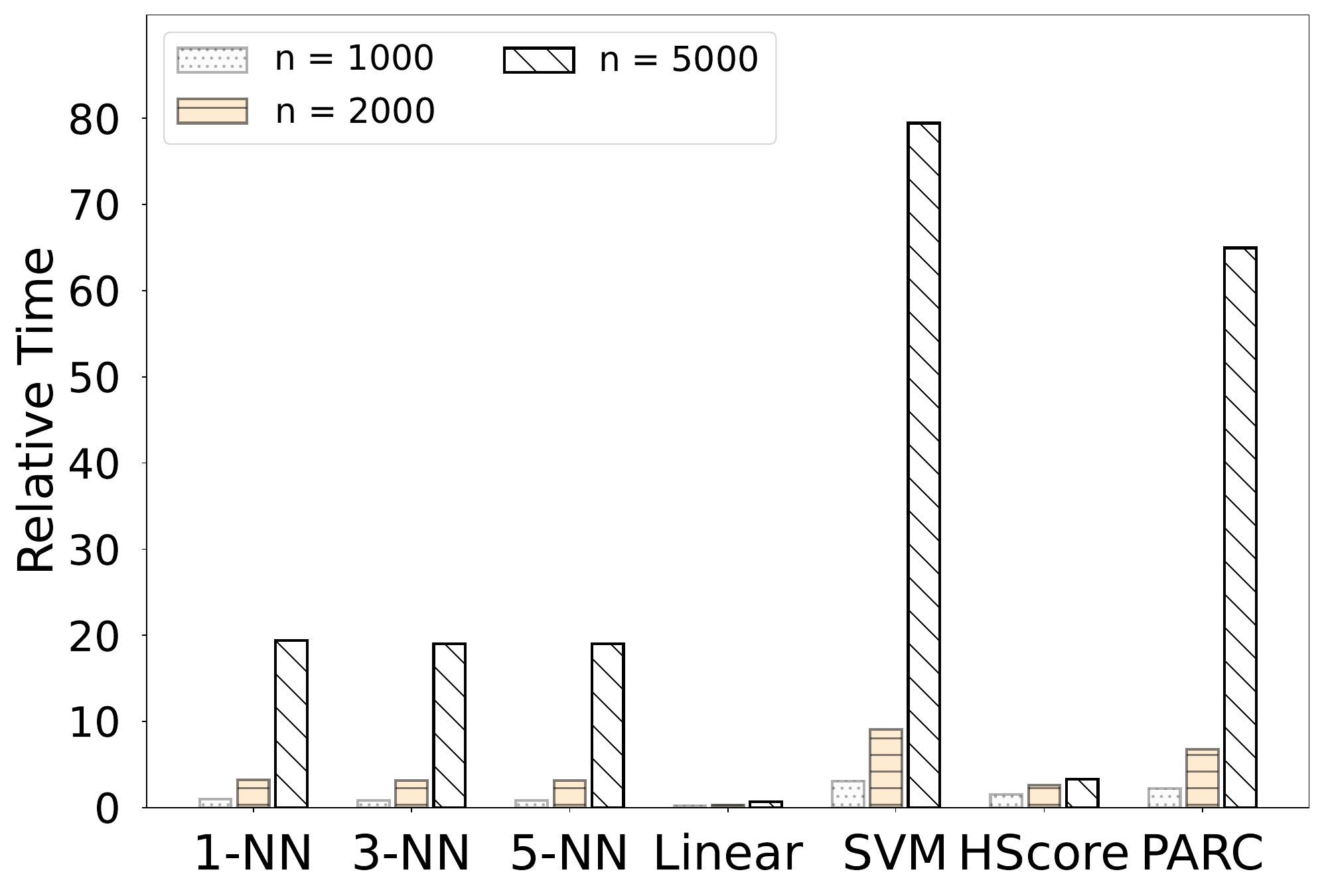}
 \caption{The time cost of learning-based model selection strategies grows in different ways by the number of probe samples (relative to 1-NN, which costs 95.37 seconds to select from 100 models using 1,000 samples)}
 \label{fig:time_analysis}
	\vspace{-1em}
\end{figure}

\subsection{Time Cost Analysis}
We assess the time cost associated with learning-based model selection strategies using 1,000, 2,000, and 5,000 probe samples. As shown in Figure~\ref{fig:time_analysis}, the time cost for each selection method increases as the sample size \(n\) grows. However, the rate of increase in time cost varies significantly across different learning strategies.
For proxy-based methods, the primary contributors to time cost are training and inference. In the case of linear methods, the number of training steps scales linearly with the sample size, while inference time remains relatively constant, resulting in a linear increase in overall time cost. 
For the \(k\)-nearest neighbors (\(k\)-NN) approach, a single inference requires comparing the current sample with all \(n/2\) samples in the space, leading to a total of \(n/2\) comparisons. As a result, the time cost increases quadratically with the sample size.
For SVM, the training process involves solving a quadratic programming problem, with computational complexity ranging from \(O(n^2)\) to \(O(n^3)\)~\cite{hearst1998support}.

For distribution-based methods, the primary time expenditure is on computing the matrices $S^f$ and $S^y$. In PARC, elements are compared pairwise, resulting in $O(n^2)$ time complexity. For the H-Score method, the calculation of the correlation matrix $\cov = (\Tilde{F}^\intercal \Tilde{F})/( F -1)$ is the main workload, where the multiplication of $\Tilde{F}$ and its transpose can be efficiently performed using a single-pass algorithm~\cite{bennett2009numerically}, necessitating $O(n)$ operations.

\section{Discussion}

\subsection{Implications of Findings}
\label{sec_implications}
This study reveals several
practical guidelines on how to select a model from a zoo of PCMs.

\mypara{Intuitive model metadata does not correlate well with downstream performance}
Our findings in three intuitive approaches have demonstrated that the optimal model cannot be readily determined by considering only metadata like the model size or dataset size. 
It remains to be investigated why a smaller model or one trained on a less extensive dataset can outperform others in downstream tasks.

\mypara{It is possible to select the model that is most fit for the target task with little computation overhead}
Our experiment results in RQ1 suggest that by leveraging the learning-based model selection method, it is possible to find the best model within a limited computation budget.
Identifying the most suitable model can yield up to a 5\% performance improvement, while intuitive methods struggle to select the best model accurately.

\mypara{The underlying principle behind learning-based model selection}
The three intuitive methods either rely solely on metadata (e.g., model size and pre-training dataset size) or resort to the computationally expensive fine-tuning process.
The learning strategies employ an intermediate-level approach, selecting models by performing a forward pass on the user-proprietary dataset, thus saving considerable effort compared to full fine-tuning, while providing promising performance.

\subsection{Threats to Validity}
\mypara{Internal Validity}
Since there is no guideline for choosing training hyperparameters for each individual model, we used the same training configuration—including learning rate, number of epochs, and batch size—for all models and selected the checkpoint that performed the best. 
Although this setting may not be fair for all models and might favor smaller ones, we find that it does not negatively impact the performance of learning-based model selection strategies. This suggests that learning-based selection strategies can effectively select models that generalize well to the target dataset, despite these minor variations.

\mypara{External Validity}
Another potential threat to validity is the size of the investigated pre-trained code models.
In this paper, although we have tried our best to collect the online PCMs from Hugging Face, several recent huge PCMs (e.g., ChatGPT~\cite{chatgpt} with up to 175 billion parameters) have been excluded due to accessibility limitations.
Additionally, our study primarily focuses on classification tasks, such as vulnerability detection. Fine-tuning PCMs for generation tasks, such as code generation or code summarization, requires significantly more computational resources due to the complexity of the sequential decoding process, compared to classification tasks that only generate an embedding vector for an input. Consequently, we leave the exploration of generation tasks and their impact on model selection strategies as future work.

\section{Related Work}

\mypara{Pre-Trained Models of Code}
Deep learning techniques are widely used to represent source code semantics~\cite{wan2024deep}, automating software engineering tasks such as 
code search~\cite{gu2018deep,wan2019multi},  summarization~\cite{hu2018deep,alon2018code2seq,wan2018improving,wang2020reinforcement,guo2022modeling}, clone detection~\cite{white2016deep,hua2020fcca}, generation~\cite{bi2024iterative}, and translation~\cite{chen2018tree}.
Code representation learning, which preserves the semantics of source code into dense or structural representation forms, is the fundamental problem~\cite{zhang2024deep}.
Inspired by the 
success of self-supervised pre-training in NLP and computer vision, 
efforts have been devoted to developing pre-trained models on large-scale code corpora for enhanced code representation~\cite{feng2020codebert,guo2020graphcodebert,li2023starcoder,touvron2023llama}.
In complementary, several works examine the interpretability~\cite{DBLP:conf/icse/WanZZSXJ22,lopez2022ast,pei2023better,shi2023towards}, transferability~\cite{DBLP:journals/corr/abs-2204-09653, tufano2023automating}, and robustness~\cite{DBLP:conf/icse/YangSH022, wang2022recode}. 
This paper examines the reusability of pre-trained code models, emphasizing model selection for fine-tuning.

\mypara{Reusability of Deep Learning Models}
Reusing 
APIs, code libraries, and models 
is common
in software development~\cite{prieto1993status, frakes2005software} due to the high expense of pre-training deep learning models~\cite{gpt-cost}.
However, model reuse involves costs and challenges 
in verifying a model's suitability for a specific task of interest~\cite{gong2023intended,you2022ranking}. 
To maximize the reuse of 
deep learning models, practices 
should adhere to guidelines of model repositories~\cite{kornblith2019better, jiang2023empirical}.
These guidelines specify the pre- and post-conditions of pre-trained models, enabling efficient model reuse without delving into model parameters
~\cite{gong2023intended, yu2024makes}. However, when guidelines are absent, methods that leverage parameters to efficiently rank available models become necessary~\cite{poth2021pre, zhao2024coding}. 
To select a model efficiently, 
existing studies 
predict the performance after fine-tuning~\cite{renggli2022NEEDLE} based on the performance of shallow proxy models~\cite{renggli2022NEEDLE, puigcerver2020scalable}, the correlation of PCM features with task labels~\cite{bolya2021PARC, bao2019HScore,zhang2024model, li2023guided}, and the performance of PCMs on a reference task~\cite{yan2020NDS}.

\section{Conclusion}

This paper discusses the efficient selection of pre-trained models for a target task.
We examine three intuitive model selection methods and find they either produce inaccuracies or require significant computational resources.
Furthermore, we explore two learning-based 
selection strategies (i.e., proxy-based and distribution-based), demonstrating their effectiveness over intuitive methods.
We assess the adaptability, scalability, and time cost of learning-based selection methods.
We hope that this work can help software engineers select the right pre-trained code model during the prototyping phase of AI-assisted code automation software development.

\mypara{Data Availability}
The code and data are available at \texttt{\url{https://github.com/CGCL-codes/naturalcc/tree/main/examples/pcm-reuse}}~\cite{wan2022naturalcc}.

\section*{Acknowledgements}
This work is supported by the Major Program (JD) of Hubei Province (Grant No. 2023BAA024), the National Natural Science Foundation of China (Grant No. 62102157), and the Experimental Technology Research Project of Huazhong University of Science and Technology (2025-2-72).

\balance
\bibliographystyle{IEEEtran}
\bibliography{ref}

\begin{thebibliography}{10}
\providecommand{\url}[1]{#1}
\csname url@samestyle\endcsname
\providecommand{\newblock}{\relax}
\providecommand{\bibinfo}[2]{#2}
\providecommand{\BIBentrySTDinterwordspacing}{\spaceskip=0pt\relax}
\providecommand{\BIBentryALTinterwordstretchfactor}{4}
\providecommand{\BIBentryALTinterwordspacing}{\spaceskip=\fontdimen2\font plus
\BIBentryALTinterwordstretchfactor\fontdimen3\font minus \fontdimen4\font\relax}
\providecommand{\BIBforeignlanguage}[2]{{%
\expandafter\ifx\csname l@#1\endcsname\relax
\typeout{** WARNING: IEEEtran.bst: No hyphenation pattern has been}%
\typeout{** loaded for the language `#1'. Using the pattern for}%
\typeout{** the default language instead.}%
\else
\language=\csname l@#1\endcsname
\fi
#2}}
\providecommand{\BIBdecl}{\relax}
\BIBdecl

\bibitem{feng2020codebert}
Z.~Feng, D.~Guo, D.~Tang, N.~Duan, X.~Feng, M.~Gong, L.~Shou, B.~Qin, T.~Liu, D.~Jiang, and M.~Zhou, ``Codebert: {A} pre-trained model for programming and natural languages,'' in \emph{Proceedings of the 2020 Conference on Empirical Methods in Natural Language Processing}, 2020, pp. 1536--1547.

\bibitem{ahmad2021unified}
W.~Ahmad, S.~Chakraborty, B.~Ray, and K.-W. Chang, ``Unified pre-training for program understanding and generation,'' in \emph{Proceedings of the 2021 Conference of the North American Chapter of the Association for Computational Linguistics: Human Language Technologies}.\hskip 1em plus 0.5em minus 0.4em\relax Online: Association for Computational Linguistics, Jun. 2021, pp. 2655--2668.

\bibitem{wang2021codet5}
Y.~Wang, W.~Wang, S.~Joty, and S.~C. Hoi, ``Codet5: Identifier-aware unified pre-trained encoder-decoder models for code understanding and generation,'' in \emph{Proceedings of the 2021 Conference on Empirical Methods in Natural Language Processing}, 2021, pp. 8696--8708.

\bibitem{nijkamp2022codegen}
E.~Nijkamp, B.~Pang, H.~Hayashi, L.~Tu, H.~Wang, Y.~Zhou, S.~Savarese, and C.~Xiong, ``Codegen: An open large language model for code with multi-turn program synthesis,'' in \emph{Proceedings of the International Conference on Learning Representations}, 2022.

\bibitem{li2023starcoder}
R.~Li, L.~B. Allal, Y.~Zi, N.~Muennighoff, D.~Kocetkov, C.~Mou, M.~Marone, C.~Akiki, J.~Li, and J.~Chim, ``Starcoder: may the source be with you!'' \emph{arXiv preprint arXiv:2305.06161}, 2023.

\bibitem{roziere2023code}
B.~Rozière, J.~Gehring, F.~Gloeckle, S.~Sootla, I.~Gat, X.~E. Tan, Y.~Adi, J.~Liu, R.~Sauvestre, T.~Remez, J.~Rapin, A.~Kozhevnikov, I.~Evtimov, J.~Bitton, M.~Bhatt, C.~C. Ferrer, A.~Grattafiori, W.~Xiong, A.~Défossez, J.~Copet, F.~Azhar, H.~Touvron, L.~Martin, N.~Usunier, T.~Scialom, and G.~Synnaeve, ``Code llama: Open foundation models for code,'' \emph{arXiv preprint arXiv:2308.12950}, 2023.

\bibitem{devlin2019bert}
J.~Devlin, M.~Chang, K.~Lee, and K.~Toutanova, ``{BERT:} pre-training of deep bidirectional transformers for language understanding,'' in \emph{Proceedings of the 2019 Conference of the North American Chapter of the Association for Computational Linguistics: Human Language Technologies}, 2019, pp. 4171--4186.

\bibitem{gpt-cost}
C.~Li, ``{Demystifying gpt-3 language model: A technical overview},'' \url{https://lambdalabs.com/blog/demystifying-gpt-3/}, 2020, [Online; accessed 1-Aug-2022].

\bibitem{huggingface}
``{Hugging Face},'' \url{https://www.huggingface.com}, 2023, [Online; accessed 1-Feb-2023].

\bibitem{tfhub}
``{Tensorflow Hub},'' \url{https://www.tensorflow.org/hub}, 2019, [Online; accessed 1-Aug-2022].

\bibitem{onnxmodelzoo}
``Onnx model zoo,'' \url{https://github.com/onnx/models}, 2023, accessed: 1-Feb-2023.

\bibitem{pytorchhub}
``Pytorch hub,'' \url{https://pytorch.org/hub/}, 2023, accessed: 1-Feb-2023.

\bibitem{gong2023intended}
L.~Gong, J.~Zhang, M.~Wei, H.~Zhang, and Z.~Huang, ``What is the intended usage context of this model? an exploratory study of pre-trained models on various model repositories,'' \emph{ACM Transactions on Software Engineering and Methodology}, vol.~32, no.~3, pp. 1--57, 2023.

\bibitem{zhang2021survey}
Y.~Zhang, P.~Ti{\v{n}}o, A.~Leonardis, and K.~Tang, ``A survey on neural network interpretability,'' \emph{IEEE Transactions on Emerging Topics in Computational Intelligence}, vol.~5, no.~5, pp. 726--742, 2021.

\bibitem{kaplan2020scaling}
J.~Kaplan, S.~McCandlish, T.~Henighan, T.~B. Brown, B.~Chess, R.~Child, S.~Gray, A.~Radford, J.~Wu, and D.~Amodei, ``Scaling laws for neural language models,'' \emph{arXiv preprint arXiv:2001.08361}, 2020.

\bibitem{renggli2022NEEDLE}
C.~Renggli, A.~S. Pinto, L.~Rimanic, J.~Puigcerver, C.~Riquelme, C.~Zhang, and M.~Lu{\v{c}}i{\'c}, ``Which model to transfer? finding the needle in the growing haystack,'' in \emph{Proceedings of the IEEE/CVF Conference on Computer Vision and Pattern Recognition}, 2022, pp. 9205--9214.

\bibitem{yan2020NDS}
X.~Yan, D.~Acuna, and S.~Fidler, ``Neural data server: A large-scale search engine for transfer learning data,'' in \emph{Proceedings of the IEEE/CVF Conference on Computer Vision and Pattern Recognition}, 2020, pp. 3893--3902.

\bibitem{puigcerver2020scalable}
J.~Puigcerver, C.~R. Ruiz, B.~Mustafa, C.~Renggli, A.~S. Pinto, S.~Gelly, D.~Keysers, and N.~Houlsby, ``Scalable transfer learning with expert models,'' in \emph{Proceedings of the International Conference on Learning Representations}, 2020.

\bibitem{bolya2021PARC}
D.~Bolya, R.~Mittapalli, and J.~Hoffman, ``Scalable diverse model selection for accessible transfer learning,'' in \emph{Advances in Neural Information Processing Systems}, vol.~34, 2021, pp. 19\,301--19\,312.

\bibitem{bao2019HScore}
Y.~Bao, Y.~Li, S.-L. Huang, L.~Zhang, L.~Zheng, A.~Zamir, and L.~Guibas, ``An information-theoretic approach to transferability in task transfer learning,'' in \emph{Proceedings of the IEEE International Conference on Image Processing}.\hskip 1em plus 0.5em minus 0.4em\relax IEEE, 2019, pp. 2309--2313.

\bibitem{husain2019codesearchnet}
H.~Husain, H.~Wu, T.~Gazit, M.~Allamanis, and M.~Brockschmidt, ``Codesearchnet challenge: Evaluating the state of semantic code search,'' \emph{arXiv preprint arXiv:1909.09436}, vol. abs/1909.09436, 2019.

\bibitem{kocetkov2022stack}
D.~Kocetkov, R.~Li, L.~B. Allal, J.~Li, C.~Mou, C.~M. Ferrandis, Y.~Jernite, M.~Mitchell, S.~Hughes, T.~Wolf, D.~Bahdanau, L.~von Werra, and H.~de~Vries, ``The stack: 3 tb of permissively licensed source code,'' \emph{arXiv preprint arXiv:2211.15533}, 2022.

\bibitem{zeng2022extensive}
Z.~Zeng, H.~Tan, H.~Zhang, J.~Li, Y.~Zhang, and L.~Zhang, ``An extensive study on pre-trained models for program understanding and generation,'' in \emph{Proceedings of the 31st ACM SIGSOFT International Symposium on Software Testing and Analysis}, 2022, pp. 39--51.

\bibitem{guo2020graphcodebert}
D.~Guo, S.~Ren, S.~Lu, Z.~Feng, D.~Tang, S.~Liu, L.~Zhou, N.~Duan, A.~Svyatkovskiy, S.~Fu, M.~Tufano, S.~K. Deng, C.~B. Clement, D.~Drain, N.~Sundaresan, J.~Yin, D.~Jiang, and M.~Zhou, ``Graphcodebert: Pre-training code representations with data flow,'' in \emph{Proceedings of the 9th International Conference on Learning Representations}, 2021.

\bibitem{lu2021codexglue}
S.~Lu, D.~Guo, S.~Ren, J.~Huang, A.~Svyatkovskiy, A.~Blanco, C.~Clement, D.~Drain, D.~Jiang, D.~Tang, G.~Li, L.~Zhou, L.~Shou, L.~Zhou, M.~Tufano, M.~Gong, M.~Zhou, N.~Duan, N.~Sundaresan, S.~K. Deng, S.~Fu, and S.~Liu, ``Codexglue: A machine learning benchmark dataset for code understanding and generation,'' \emph{arXiv preprint arXiv:2102.04664}, 2021.

\bibitem{zheng2023codegeex}
Q.~Zheng, X.~Xia, X.~Zou, Y.~Dong, S.~Wang, Y.~Xue, L.~Shen, Z.~Wang, A.~Wang, Y.~Li, T.~Su, Z.~Yang, and J.~Tang, ``Codegeex: A pre-trained model for code generation with multilingual benchmarking on humaneval-x,'' in \emph{Proceedings of the 29th ACM SIGKDD Conference on Knowledge Discovery and Data Mining}.\hskip 1em plus 0.5em minus 0.4em\relax New York, NY, USA: Association for Computing Machinery, 2023, p. 5673–5684.

\bibitem{li2022competition}
Y.~Li, D.~Choi, J.~Chung, N.~Kushman, J.~Schrittwieser, R.~Leblond, T.~Eccles, J.~Keeling, F.~Gimeno, A.~D. Lago, T.~Hubert, P.~Choy, C.~de~Masson~d’Autume, I.~Babuschkin, X.~Chen, P.-S. Huang, J.~Welbl, S.~Gowal, A.~Cherepanov, J.~Molloy, D.~J. Mankowitz, E.~S. Robson, P.~Kohli, N.~de~Freitas, K.~Kavukcuoglu, and O.~Vinyals, ``Competition-level code generation with alphacode,'' \emph{Science}, vol. 378, no. 6624, pp. 1092--1097, 2022.

\bibitem{Clark2020ELECTRA}
K.~Clark, M.-T. Luong, Q.~V. Le, and C.~D. Manning, ``Electra: Pre-training text encoders as discriminators rather than generators,'' in \emph{Proceedings of International Conference on Learning Representations}, 2020.

\bibitem{lewis2020bart}
M.~Lewis, Y.~Liu, N.~Goyal, M.~Ghazvininejad, A.~Mohamed, O.~Levy, V.~Stoyanov, and L.~Zettlemoyer, ``{BART}: Denoising sequence-to-sequence pre-training for natural language generation, translation, and comprehension,'' in \emph{Proceedings of the 58th Annual Meeting of the Association for Computational Linguistics}, Jul. 2020, pp. 7871--7880.

\bibitem{raffel2020exploring}
C.~Raffel, N.~Shazeer, A.~Roberts, K.~Lee, S.~Narang, M.~Matena, Y.~Zhou, W.~Li, and P.~J. Liu, ``Exploring the limits of transfer learning with a unified text-to-text transformer,'' \emph{Journal of Machine Learning Research}, vol.~21, no. 140, pp. 1--67, 2020.

\bibitem{zhou2019devign}
Y.~Zhou, S.~Liu, J.~Siow, X.~Du, and Y.~Liu, ``Devign: Effective vulnerability identification by learning comprehensive program semantics via graph neural networks,'' in \emph{Advances in Neural Information Processing Systems}, H.~Wallach, H.~Larochelle, A.~Beygelzimer, F.~d\textquotesingle Alch\'{e}-Buc, E.~Fox, and R.~Garnett, Eds., vol.~32.\hskip 1em plus 0.5em minus 0.4em\relax Curran Associates, Inc., 2019.

\bibitem{mou2016convolutional}
L.~Mou, G.~Li, L.~Zhang, T.~Wang, and Z.~Jin, ``Convolutional neural networks over tree structures for programming language processing,'' in \emph{Proceedings of the AAAI Conference on Artificial Intelligence}, vol.~30, no.~1, 2016.

\bibitem{alreshedy2018scc}
K.~Alreshedy, D.~Dharmaretnam, D.~M. German, V.~Srinivasan, and T.~A. Gulliver, ``Scc: Automatic classification of code snippets,'' in \emph{Proceedings of the IEEE International Working Conference on Source Code Analysis and Manipulation}, 2018, pp. 203--208.

\bibitem{zhou2023codebertscore}
S.~Zhou, U.~Alon, S.~Agarwal, and G.~Neubig, ``{C}ode{BERTS}core: Evaluating code generation with pretrained models of code,'' in \emph{Proceedings of the 2023 Conference on Empirical Methods in Natural Language Processing}, H.~Bouamor, J.~Pino, and K.~Bali, Eds.\hskip 1em plus 0.5em minus 0.4em\relax Singapore: Association for Computational Linguistics, Dec. 2023, pp. 13\,921--13\,937.

\bibitem{codeparrot}
``{Code Parrot},'' \url{https://huggingface.co/datasets/codeparrot/codeparrot-clean}, 2023, [Online; accessed 1-Feb-2023].

\bibitem{DBLP:journals/corr/abs-2204-09653}
F.~Chen, F.~H. Fard, D.~Lo, and T.~Bryksin, ``On the transferability of pre-trained language models for low-resource programming languages,'' in \emph{Proceedings of the 30th IEEE/ACM International Conference on Program Comprehension}, 2022, pp. 401--412.

\bibitem{peng2021could}
D.~Peng, S.~Zheng, Y.~Li, G.~Ke, D.~He, and T.-Y. Liu, ``How could neural networks understand programs?'' in \emph{Proceedings of the International Conference on Machine Learning}.\hskip 1em plus 0.5em minus 0.4em\relax PMLR, 2021, pp. 8476--8486.

\bibitem{jarvelin2002cumulated}
K.~J\"{a}rvelin and J.~Kek\"{a}l\"{a}inen, ``Cumulated gain-based evaluation of ir techniques,'' \emph{ACM Transactions on Information Systems (TOIS)}, vol.~20, no.~4, p. 422–446, oct 2002.

\bibitem{li2021NLEEP}
Y.~Li, X.~Jia, R.~Sang, Y.~Zhu, B.~Green, L.~Wang, and B.~Gong, ``Ranking neural checkpoints,'' in \emph{Proceedings of the IEEE/CVF Conference on Computer Vision and Pattern Recognition}, 2021, pp. 2663--2673.

\bibitem{agostinelli2022STABLE}
A.~Agostinelli, M.~P{\'a}ndy, J.~Uijlings, T.~Mensink, and V.~Ferrari, ``How stable are transferability metrics evaluations?'' in \emph{Proceedings of the European Conference on Computer Vision}.\hskip 1em plus 0.5em minus 0.4em\relax Springer, 2022, pp. 303--321.

\bibitem{hearst1998support}
M.~A. Hearst, S.~T. Dumais, E.~Osuna, J.~Platt, and B.~Scholkopf, ``Support vector machines,'' \emph{IEEE Intelligent Systems and Their Applications}, vol.~13, no.~4, pp. 18--28, 1998.

\bibitem{bennett2009numerically}
J.~Bennett, R.~Grout, P.~P{\'e}bay, D.~Roe, and D.~Thompson, ``Numerically stable, single-pass, parallel statistics algorithms,'' in \emph{Proceedings of the IEEE International Conference on Cluster Computing and Workshops}.\hskip 1em plus 0.5em minus 0.4em\relax IEEE, 2009, pp. 1--8.

\bibitem{chatgpt}
``{chatgpt},'' \url{http://chat.openai.com}, 2023, [Online; accessed 1-Feb-2023].

\bibitem{wan2024deep}
Y.~Wan, Z.~Bi, Y.~He, J.~Zhang, H.~Zhang, Y.~Sui, G.~Xu, H.~Jin, and P.~Yu, ``Deep learning for code intelligence: Survey, benchmark and toolkit,'' \emph{ACM Computing Surveys}, 2024.

\bibitem{gu2018deep}
X.~Gu, H.~Zhang, and S.~Kim, ``Deep code search,'' in \emph{Proceedings of 40th International Conference on Software Engineering}, 2018, pp. 933--944.

\bibitem{wan2019multi}
Y.~Wan, J.~Shu, Y.~Sui, G.~Xu, Z.~Zhao, J.~Wu, and P.~S. Yu, ``Multi-modal attention network learning for semantic source code retrieval,'' in \emph{Proceedings of 34th {IEEE/ACM} International Conference on Automated Software Engineering}.\hskip 1em plus 0.5em minus 0.4em\relax {IEEE}, 2019, pp. 13--25.

\bibitem{hu2018deep}
X.~Hu, G.~Li, X.~Xia, D.~Lo, and Z.~Jin, ``Deep code comment generation,'' in \emph{Proceedings of the IEEE/ACM 26th International Conference on Program Comprehension}.\hskip 1em plus 0.5em minus 0.4em\relax IEEE, 2018, pp. 200--20\,010.

\bibitem{alon2018code2seq}
U.~Alon, S.~Brody, O.~Levy, and E.~Yahav, ``code2seq: Generating sequences from structured representations of code,'' in \emph{Proceedings of the International Conference on Learning Representations}, 2018.

\bibitem{wan2018improving}
Y.~Wan, Z.~Zhao, M.~Yang, G.~Xu, H.~Ying, J.~Wu, and P.~S. Yu, ``Improving automatic source code summarization via deep reinforcement learning,'' in \emph{Proceedings of the 33rd {ACM/IEEE} International Conference on Automated Software Engineering}.\hskip 1em plus 0.5em minus 0.4em\relax {ACM}, 2018, pp. 397--407.

\bibitem{wang2020reinforcement}
W.~Wang, Y.~Zhang, Y.~Sui, Y.~Wan, Z.~Zhao, J.~Wu, S.~Y. Philip, and G.~Xu, ``Reinforcement-learning-guided source code summarization using hierarchical attention,'' \emph{IEEE Transactions on software Engineering}, vol.~48, no.~1, pp. 102--119, 2020.

\bibitem{guo2022modeling}
J.~Guo, J.~Liu, Y.~Wan, L.~Li, and P.~Zhou, ``Modeling hierarchical syntax structure with triplet position for source code summarization,'' in \emph{Proceedings of the 60th Annual Meeting of the Association for Computational Linguistics (Volume 1: Long Papers)}, 2022, pp. 486--500.

\bibitem{white2016deep}
M.~White, M.~Tufano, C.~Vendome, and D.~Poshyvanyk, ``Deep learning code fragments for code clone detection,'' in \emph{Proceedings of the 31st IEEE/ACM International Conference on Automated Software Engineering}.\hskip 1em plus 0.5em minus 0.4em\relax ACM, 2016, pp. 87--98.

\bibitem{hua2020fcca}
W.~Hua, Y.~Sui, Y.~Wan, G.~Liu, and G.~Xu, ``Fcca: Hybrid code representation for functional clone detection using attention networks,'' \emph{IEEE Transactions on Reliability}, vol.~70, no.~1, pp. 304--318, 2020.

\bibitem{bi2024iterative}
Z.~Bi, Y.~Wan, Z.~Wang, H.~Zhang, B.~Guan, F.~Lu, Z.~Zhang, Y.~Sui, H.~Jin, and X.~Shi, ``Iterative refinement of project-level code context for precise code generation with compiler feedback,'' in \emph{Proceedings of the 62nd Annual Meeting of the Association for Computational Linguistics (ACL 2024)}.\hskip 1em plus 0.5em minus 0.4em\relax Bangkok, Thailand: Association for Computational Linguistics, Aug. 2024, pp. 2336--2353.

\bibitem{chen2018tree}
X.~Chen, C.~Liu, and D.~Song, ``Tree-to-tree neural networks for program translation,'' in \emph{Advances in Neural Information Processing Systems}, 2018, pp. 2552--2562.

\bibitem{zhang2024deep}
H.~Zhang, K.~Zhang, Z.~Li, J.~Li, Y.~Li, Y.~Zhao, Y.~Zhu, F.~Liu, G.~Li, and Z.~Jin, ``Deep learning for code generation: A survey,'' \emph{SCIENCE CHINA Information Sciences}, 2024.

\bibitem{touvron2023llama}
H.~Touvron, T.~Lavril, G.~Izacard, X.~Martinet, M.-A. Lachaux, T.~Lacroix, B.~Rozière, N.~Goyal, E.~Hambro, F.~Azhar, A.~Rodriguez, A.~Joulin, E.~Grave, and G.~Lample, ``Llama: Open and efficient foundation language models,'' \emph{arXiv preprint arXiv:2302.13971}, 2023.

\bibitem{DBLP:conf/icse/WanZZSXJ22}
Y.~Wan, W.~Zhao, H.~Zhang, Y.~Sui, G.~Xu, and H.~Jin, ``What do they capture? - {A} structural analysis of pre-trained language models for source code,'' in \emph{Proceedings of the {IEEE/ACM} 44th International Conference on Software Engineering}.\hskip 1em plus 0.5em minus 0.4em\relax {ACM}, 2022, pp. 2377--2388.

\bibitem{lopez2022ast}
J.~A. Hern\'{a}ndez~L\'{o}pez, M.~Weyssow, J.~S. Cuadrado, and H.~Sahraoui, ``Ast-probe: Recovering abstract syntax trees from hidden representations of pre-trained language models,'' in \emph{Proceedings of the 37th IEEE/ACM International Conference on Automated Software Engineering}, ser. ASE '22.\hskip 1em plus 0.5em minus 0.4em\relax New York, NY, USA: Association for Computing Machinery, 2023.

\bibitem{pei2023better}
H.~Pei, J.~Zhao, L.~Lausen, S.~Zha, and G.~Karypis, ``Better context makes better code language models: a case study on function call argument completion,'' in \emph{Proceedings of the Thirty-Seventh AAAI Conference on Artificial Intelligence}.\hskip 1em plus 0.5em minus 0.4em\relax AAAI Press, 2023.

\bibitem{shi2023towards}
E.~Shi, Y.~Wang, H.~Zhang, L.~Du, S.~Han, D.~Zhang, and H.~Sun, ``Towards efficient fine-tuning of pre-trained code models: An experimental study and beyond,'' in \emph{Proceedings of the 32nd ACM SIGSOFT International Symposium on Software Testing and Analysis}, 2023, pp. 39--51.

\bibitem{tufano2023automating}
R.~Tufano, L.~Pascarella, and G.~Bavota, ``Automating code-related tasks through transformers: The impact of pre-training,'' in \emph{Proceedings of the IEEE/ACM 45th International Conference on Software Engineering (ICSE)}.\hskip 1em plus 0.5em minus 0.4em\relax IEEE, 2023, pp. 2425--2437.

\bibitem{DBLP:conf/icse/YangSH022}
Z.~Yang, J.~Shi, J.~He, and D.~Lo, ``Natural attack for pre-trained models of code,'' in \emph{Proceedings of the 44th {IEEE/ACM} 44th International Conference on Software Engineering, {ICSE} 2022, Pittsburgh, PA, USA, May 25-27, 2022}.\hskip 1em plus 0.5em minus 0.4em\relax {ACM}, 2022, pp. 1482--1493.

\bibitem{wang2022recode}
S.~Wang, Z.~Li, H.~Qian, C.~Yang, Z.~Wang, M.~Shang, V.~Kumar, S.~Tan, B.~Ray, P.~Bhatia, R.~Nallapati, M.~K. Ramanathan, D.~Roth, and B.~Xiang, ``{R}e{C}ode: Robustness evaluation of code generation models,'' in \emph{Proceedings of the 61st Annual Meeting of the Association for Computational Linguistics}, Jul. 2023, pp. 13\,818--13\,843.

\bibitem{prieto1993status}
R.~Prieto-Diaz, ``Status report: Software reusability,'' \emph{IEEE Software}, vol.~10, no.~3, pp. 61--66, 1993.

\bibitem{frakes2005software}
W.~B. Frakes and K.~Kang, ``Software reuse research: Status and future,'' \emph{IEEE Transactions on Software Engineering}, vol.~31, no.~7, pp. 529--536, 2005.

\bibitem{you2022ranking}
K.~You, Y.~Liu, Z.~Zhang, J.~Wang, M.~I. Jordan, and M.~Long, ``Ranking and tuning pre-trained models: A new paradigm for exploiting model hubs,'' \emph{Journal of Machine Learning Research}, vol.~23, no. 209, pp. 1--47, 2022.

\bibitem{kornblith2019better}
S.~Kornblith, J.~Shlens, and Q.~V. Le, ``Do better imagenet models transfer better?'' in \emph{Proceedings of the IEEE/CVF Conference on Computer Vision and Pattern Recognition}, 2019, pp. 2661--2671.

\bibitem{jiang2023empirical}
W.~Jiang, N.~Synovic, M.~Hyatt, T.~R. Schorlemmer, R.~Sethi, Y.-H. Lu, G.~K. Thiruvathukal, and J.~C. Davis, ``An empirical study of pre-trained model reuse in the hugging face deep learning model registry,'' in \emph{Proceedings of the 2023 IEEE/ACM 45th International Conference on Software Engineering}.\hskip 1em plus 0.5em minus 0.4em\relax IEEE, 2023, pp. 2463--2475.

\bibitem{yu2024makes}
X.~Yu, Z.~Zhang, F.~Niu, X.~Hu, X.~Xia, and J.~Grundy, ``What makes a high-quality training dataset for large language models: A practitioners' perspective,'' in \emph{Proceedings of the 39th IEEE/ACM International Conference on Automated Software Engineering}, 2024, pp. 656--668.

\bibitem{poth2021pre}
C.~Poth, J.~Pfeiffer, A.~R{\"u}ckl{\'e}, and I.~Gurevych, ``What to pre-train on? efficient intermediate task selection,'' \emph{arXiv preprint arXiv:2104.08247}, 2021.

\bibitem{zhao2024coding}
Y.~Zhao, L.~Gong, Z.~Huang, Y.~Wang, M.~Wei, and F.~Wu, ``Coding-ptms: How to find optimal code pre-trained models for code embedding in vulnerability detection?'' in \emph{Proceedings of the 39th IEEE/ACM International Conference on Automated Software Engineering}, 2024, pp. 1732--1744.

\bibitem{zhang2024model}
Y.-K. Zhang, T.-J. Huang, Y.-X. Ding, D.-C. Zhan, and H.-J. Ye, ``Model spider: Learning to rank pre-trained models efficiently,'' in \emph{Advances in Neural Information Processing Systems}, vol.~36, 2024.

\bibitem{li2023guided}
H.~Li, C.~Fowlkes, H.~Yang, O.~Dabeer, Z.~Tu, and S.~Soatto, ``Guided recommendation for model fine-tuning,'' in \emph{Proceedings of the IEEE/CVF Conference on Computer Vision and Pattern Recognition}, 2023, pp. 3633--3642.

\bibitem{wan2022naturalcc}
Y.~Wan, Y.~He, Z.~Bi, J.~Zhang, Y.~Sui, H.~Zhang, K.~Hashimoto, H.~Jin, G.~Xu, C.~Xiong, and P.~S. Yu, ``Naturalcc: An open-source toolkit for code intelligence,'' in \emph{Proceedings of 44th International Conference on Software Engineering, Companion Volume}.\hskip 1em plus 0.5em minus 0.4em\relax {ACM}, 2022.

\end{thebibliography}

\end{document}